\documentclass[12pt,final,onecolumn]{IEEEtran}
\ifCLASSINFOpdf
\else
\fi
\usepackage{makeidx}
\usepackage{ifpdf}
\usepackage{dsfont}
\usepackage{mathrsfs}
\usepackage{eso-pic}

\newcommand{\bitem}{\begin{itemize}}
\newcommand{\eitem}{\end{itemize}}
\newcommand{\beq}{\begin{equation}}
\newcommand{\eeq}{\end{equation}}
\newcommand{\beqa}{\begin{array}}
\newcommand{\eeqa}{\end{array}}
\newcommand{\beqar}{\begin{eqnarray}}
\newcommand{\eeqar}{\end{eqnarray}}
\newcommand{\bef}{\begin{figure}}
\newcommand{\enf}{\end{figure}}
\newcommand{\btab}{\begin{tabular}}
\newcommand{\etab}{\end{tabular}}
\newcommand{\btable}{\begin{table}}
\newcommand{\etable}{\end{table}}
\newcommand{\benum}{\begin{enumerate}}
\newcommand{\eenum}{\end{enumerate}}
\newcommand{\bdis}{\begin{displaymath}}
\newcommand{\edis}{\end{displaymath}}

\usepackage{color}

\newcommand{\squishlist}{
   \begin{list}{$\bullet$}
    { \setlength{\itemsep}{0pt}      \setlength{\parsep}{3pt}
      \setlength{\topsep}{3pt}       \setlength{\partopsep}{0pt}
      \setlength{\leftmargin}{1.5em} \setlength{\labelwidth}{1em}
      \setlength{\labelsep}{0.5em} } }
\newcommand{\squishlisttwo}{
   \begin{list}{$\bullet$}
    { \setlength{\itemsep}{0pt}    \setlength{\parsep}{0pt}
      \setlength{\topsep}{0pt}     \setlength{\partopsep}{0pt}
      \setlength{\leftmargin}{2em} \setlength{\labelwidth}{1.5em}
      \setlength{\labelsep}{0.5em} } }
\newcommand{\squishend}{
    \end{list}  }

\usepackage{cite}
\usepackage[dvips]{graphicx}
\usepackage[cmex10]{amsmath}
\usepackage{algorithmic}
\usepackage{array}
\usepackage{mdwmath}
\usepackage{mdwtab}
\usepackage{eqparbox}
\usepackage[tight,footnotesize]{subfigure}
\usepackage{fixltx2e}
\usepackage{url}

\usepackage{amssymb}

\usepackage{psfrag}
\usepackage{subfigure}
\usepackage{algorithmic}
\usepackage{algorithm}
\usepackage{xspace}

\usepackage{times}

\usepackage[innercaption]{sidecap}
\usepackage{graphicx}
\usepackage{bm}
\makeindex
\usepackage{setspace}

\makeatletter
\def\ifundefined{\@ifundefined}
\makeatother
\begin{document}
\title{Modeling IEEE 802.15.4 Networks over Fading Channels}
\author{Piergiuseppe Di Marco$^{\dag}$, Carlo Fischione$^{\dag}$, \\ Fortunato Santucci$^{\ddag}$, and Karl Henrik Johansson$^{\dag}$
\thanks{Part of this work appears in the proceedings of the IEEE International Conference on Communications 2013. 
$\dag$ ACCESS Linnaeus Center, Electrical
Engineering, Royal Institute of Technology, Stockholm, Sweden.
E-mail: \texttt{\{pidm|carlofi|kallej\}@kth.se}. $\ddag$
Centre of Excellence DEWS and Dept. DISIM, University of L'Aquila, L'Aquila, Italy.
E-mail: \texttt{fortunato.santucci@univaq.it}. The authors acknowledge
the support of the Swedish Foundation for Strategic Research, the EU projects Hydrobionets and Hycon2, and the PRIN Greta project.}}

\maketitle
\begin{abstract}
Although the performance of the medium
access control (MAC) of the IEEE 802.15.4 has been
investigated under the assumption of ideal wireless channel, the
understanding of the cross-layer
dynamics between MAC and physical layer is an open problem when the wireless channel exhibits path loss, multi-path fading, and
shadowing. The analysis of MAC and wireless channel interaction
 is essential for consistent performance prediction, correct design and
optimization of the protocols. In this paper, a
novel approach to analytical modeling of these interactions is
proposed. The analysis considers simultaneously a composite channel
fading, interference generated by multiple terminals, the effects induced by hidden terminals, and the MAC
reduced carrier sensing capabilities. 
Depending on the MAC parameters and physical layer thresholds, it is shown that the MAC performance indicators over fading channels can be
far from those derived under ideal channel assumptions. 
As novel results, we show to what extent the presence of fading may be beneficial for the overall network performance by reducing the multiple access interference,
 and how this can be used to drive joint selection of MAC and physical layer parameters.
\end{abstract}
\IEEEpeerreviewmaketitle

\begin{keywords}
IEEE~802.15.4, WSN, Medium Access Control,
Fading Channel, Interference, Multi-hop.
\end{keywords}

\section{Introduction}

The development of wireless sensor network (WSN) systems relies heavily on understanding the behavior of
underlying communication mechanisms. When sensors and actuators are
integrated within the physical world with large-scale and dense deployments,
potential mobility of nodes, obstructions to propagation,
fading of the wireless channel and multi-hop networking must be
carefully addressed to offer reliable services. In fact, wireless interfaces can represent bottlenecks as they may not provide
links as solid as required by applications in terms of
reliability, delay, and energy.

There is consensus that the protocols for physical layer
and medium access control (MAC) for low data rate and low power
applications in the future will be based on the
flexible IEEE 802.15.4 standard with its numerous
variants~\cite{ieee802154}. That standard has been indeed adopted with
some modifications also by a number of other protocol stacks, including
ZigBee, WirelessHART, ISA-100~\cite{willig08}. It is already being used for applications
in industrial control, home automation, health care, and
smart grids. Nevertheless, there is not yet a
clear understanding of the achievable performance
of the IEEE 802.15.4 protocol stack, with the consequent inability to adapt the
communication performance (e.g., through cross-layer optimization) to meet challenging quality of service
requirements.

The IEEE 802.15.4 MAC layer has received much attention, with focus on
performance characterization in terms of reliability (i.e., successful packet reception probability), packet delay, throughput, and energy consumption.
Some initial works, such as~\cite{Zheng04}, are based on Monte
Carlo simulations. More recent investigations have attempted to
model the protocol performance by theoretical analysis
for single hop
networks~\cite{Misic06_beacon,Sinem08,Park_mass09,Jung09,he09,Buratti_TVT,Faridi10}.
These analytical studies are based on extensions of the Markov
chain model originally proposed by Bianchi for the IEEE 802.11 MAC
protocol~\cite{Bianchi00} and assume ideal channel conditions. 

The main limitation of the existing studies in literature is that MAC and physical layers analysis are investigated independently. 
In~\cite{jointphymac}, modeling of packet losses
due to channel fading have been introduced into the homogeneous
Markov chain developed for the IEEE~802.15.4 MAC setup presented
in~\cite{Park_mass09}. However, fading is considered only for single
packet transmission attempts, the effect of contention and multiple access
interference is neglected, and the analysis is neither validated
by simulations nor by experiments.
In~\cite{infocom05} the optimal carrier sensing range
is derived to maximize the throughput for IEEE 802.11 networks;
however, statistical modeling of wireless fading has not
been considered, but
a two-ray ground radio propagation
model is used.
Recent studies have investigated the
performance of multiple access networks in terms of multiple
access interference and capture effect for IEEE 802.11 MAC in~\cite{802.11cap,Hoang08,Leo13,Sutton13} and for IEEE 802.15.4 MAC in~\cite{buratti_capture}.
However, the models in~\cite{802.11cap,Hoang08,Leo13,buratti_capture} are limited to homogeneous networks (same statistical model for every node) with homogeneous traffic and uniform random deployment. Heterogeneous traffic conditions are discussed in~\cite{Sutton13}, by assuming two classes of traffics.
It is worthwhile mentioning that the models in~\cite{Leo13,Sutton13} represent the state of the art for the analysis of the IEEE 802.11 MAC over fading channels. Nevertheless, they consider only multi-path fading and the statistics are derived under the assumption of perfect power control and perfect carrier sensing. 
The model in~\cite{buratti_capture} assumes that nodes are synchronized and a single packet transmission for each node is considered.
Thus, the number of contending nodes in transmission is known at the beginning of the superframe.
We consider instead a setup with asynchronous Poisson traffic generation, which is more general.
Moreover, in~\cite{buratti_capture}  the channel is characterized on a distance-based model, and the effect of aggregated shadowing and multi-path components has not been considered, 
while it is known that it has a crucial impact on the performance of packet access mechanisms~\cite{Iver09}.

In all the aforementioned studies, the probability of fading and capture are evaluated in terms of average effects of the network on the tagged node. 
There is actually a closer interaction between MAC and physical channel.
For instance, a bad channel condition during the channel sensing procedure can determine more packet transmissions for the tagged node with respect to the ideal case, therefore more potential collisions. However, a bad channel condition for the contenders can imply a higher probability of success for the tagged node. 
These situations cannot be modeled by using existing analytical studies for homogeneous IEEE 802.15.4 networks (e.g.,~\cite{buratti_capture}). 
Similarly, the interactions between MAC and physical channel cannot be predicted by existing models for heterogeneous IEEE 802.15.4 networks (e.g.,~\cite{PG_TVT}), since only ideal channel conditions are considered. Finally, we remark that the combined effects of fading and multiple access interference cannot be distinguished just by mean of experimental evaluations~\cite{buratti_capture}.

In this paper we propose a novel analytical model that captures the
cross-layer interactions of IEEE 802.15.4 MAC and physical layer over interference-limited wireless channels with composite fading models.
The main original contributions are as follows.
\begin{itemize}
\item We propose a general modeling approach for
characterization of the MAC performance with heterogeneous
network conditions, a composite Nakagami-lognormal channel, explicit interference behaviors and cross-layer interactions. 
\item Based on the new model, we determine
the impact of fading conditions on the MAC performance under
various settings for traffic, inter-node distances, carrier sensing range,
and signal-to-(interference plus noise)-ratio (SINR). We show how existing models of the MAC from the literature may give unsatisfactory or inadequate predictions for performance indicators in fading channels.
\item We discuss system configurations in which a certain severity of the fading may be beneficial for overall network performance. Based on the new model, it is then possible to derive optimization guidelines for the overall network performance, by leveraging on the MAC-physical layer interactions.
\end{itemize}

To determine the network operating point and the performance
indicators in terms of reliability, delay, and energy
consumption for single-hop and multi-hop topologies, a moment matching
approximation for the linear combination of lognormal random
variables based on~\cite{pratesiTWC} and \cite{fischioneTCOMM} is adopted in
order to build a Markov chain model of the MAC mechanism that embeds the physical layer behavior.
The challenging part of the new analytical setup proposed
in this paper is to model the complex
interaction between the MAC protocol and the wireless
channel with explicit description of the dependence on several
topological parameters and network dynamics. For example, we include failures of the
channel sensing mechanism and the presence of hidden terminals, namely
nodes that are in the communication range of the destination
but cannot be listened by the transmitter. Whether two wireless nodes can communicate
with each other depends on their relative distance, the
transmission power, the wireless propagation characteristics
and interference caused by concurrent transmissions on the same
radio channel: the higher the SINR is,
the higher the probability that packets can be successfully
received. The number of concurrent transmissions depends on the traffic and the MAC parameters.
To the best of our knowledge, this is the first paper that account for statistical fluctuations of the
SINR in the Markov chain model of the IEEE 802.15.4 MAC.

The remainder of the paper is organized as follows. In
Section~\ref{sec:sysmod}, we introduce the network model. In
Section~\ref{sec:802.15.4}, we derive an analytical model of
IEEE~802.15.4 MAC over fading channels. In Section~\ref{sec:reliab}, reliability, delay, and energy consumption are derived. The accuracy of the model
is evaluated in
Section~\ref{sec:simulations}, along with a detailed analysis of performance indexes with various parameter settings. Section~\ref{sec:conclusions} concludes
the paper and prospects our future work.

\section{Network Model}\label{sec:sysmod}

\begin{figure}[t] \centering
\includegraphics[width=0.75\textwidth]{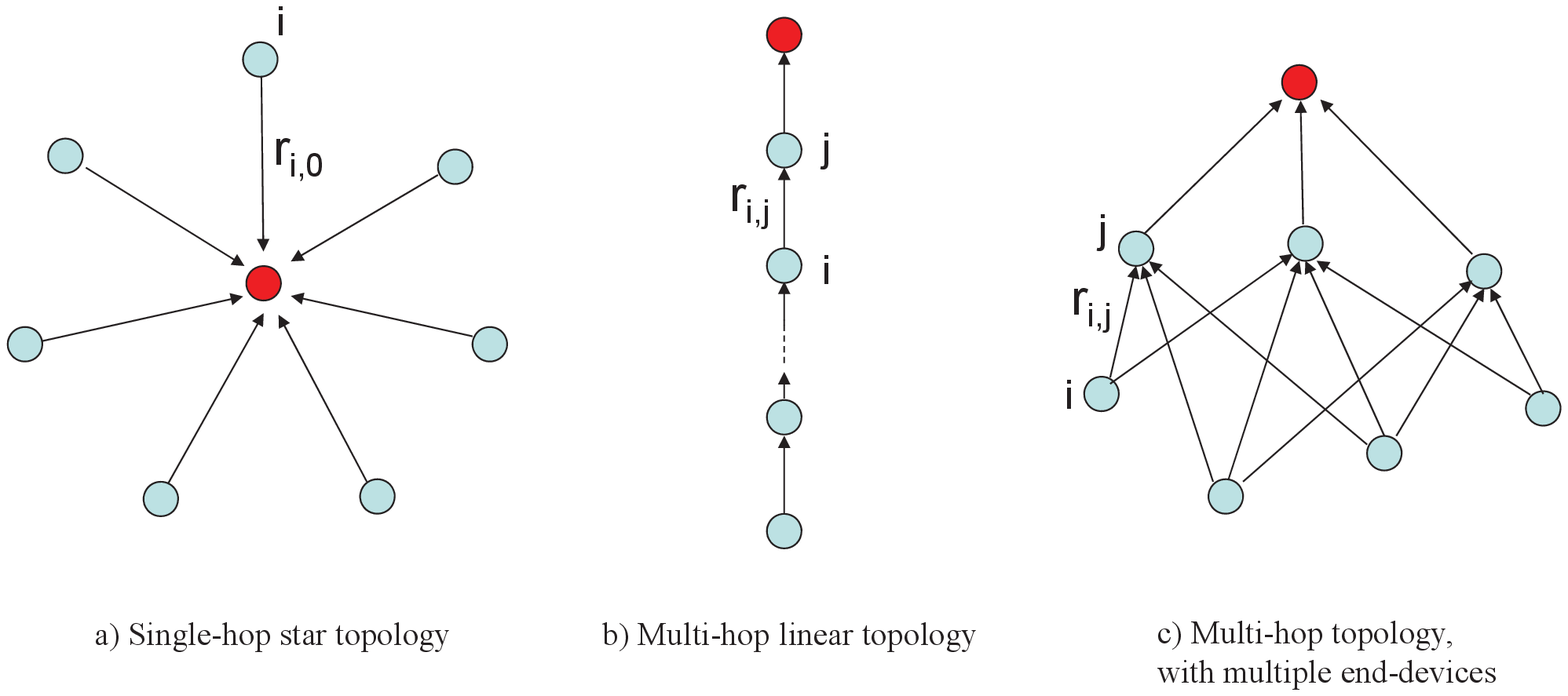}
\caption{Example of topologies: single-hop star topology (on the
left), multi-hop linear topology (in the center) and multi-hop topology with multiple end-devices (on the right).}
\label{fig:topology}\end{figure}

We illustrate the network model by considering the three topologies
sketched in Fig.~\ref{fig:topology}. Nevertheless, the analytical
results that we derive in this paper are  applicable to
any fixed topology.

The topology in Fig.~\ref{fig:topology}a) refers to a
single-hop (star) network, where node $i$ is deployed at distance $r_{i,0}$
from the root node at the center, and where nodes forward their packets
with single-hop communication to the root node. The topology
in Fig.~\ref{fig:topology}b) is a multi-hop linear topology,
where every node generates and forwards traffic to the root node
by multi-hop communication. The distance between two adjacent
nodes is denoted as $r_{i,j}$. In Fig.~\ref{fig:topology}c), we illustrate a multi-hop topology
with multiple end-devices that generate and forward traffic according to an uplink routing policy
to the root node.

Consider node $i$ that is transmitting a packet with transmission
power $P_{\mathrm{tx},i}$. We consider an inverse power
model of the link gain, and include shadowing and multi-path fading as well. The received power at node $j$, which is
located at a distance $r_{i,j}$, is then expressed as follows
\begin{align} \label{eq:prx}
P_{\mathrm{rx},i,j}=\frac{c_0 P_{\mathrm{tx},i}}{r^{k}_{i,j}} f_i \exp(y_i)\,.
\end{align}
The constant $c_0$ represents the power gain at the reference
distance $1$~m, and it can account for specific propagation environments and
parameters, e.g., carrier
frequency and antennas. In the operating conditions for
IEEE~802.15.4 networks, the inverse of $c_0$ (i.e., the path loss at the reference distance) is in the range
$40-60$~dB~\cite{ieee802154}. The exponent $k$ is called path loss
exponent, and varies according to the propagation environment in
the range $2-4$.
The factor $f_i$ models a frequency-flat channel fading due to multi-path propagation, which we assume to follow a Nakagami distribution with parameter $\kappa\geq 0.5$ and p.d.f.
$$p_{f_i}(z)= \kappa^\kappa \frac{(z)^{\kappa-1}}{\Gamma(\kappa)} \exp(-\kappa z), $$
where $\Gamma(\kappa)$ is the standard Gamma function $\Gamma(\kappa)=\int_0^\infty \exp(-x) x^{\kappa-1} dx$. We consider the Nakagami distribution since it is a general statistical model and it captures fading environments with various degrees of severity, including Rayleigh and Rice environments.
A lognormal random component models the shadowing
effects due to obstacles, with $y_i \sim \mathcal{N}(0,\sigma_i^2)$. The
standard deviation $\sigma_i$ is called spread factor of the
shadowing. These assumptions are accurate for IEEE~802.15.4 in a home or urban
environment where devices may not be in
visibility.  

In the rest of the paper, we use the index $l$ to denote a link, where $i$ is the transmitting node and $j$ is the receiving node. We use the double indices $(i,j)$ for variables that depend on a generic pair of nodes in the network.
In the following section, a generalized model of a heterogeneous
network using unslotted IEEE~802.15.4 MAC over multi-path
fading channels is proposed.

%

\section{IEEE~802.15.4 MAC and PHY Layer Model} \label{sec:802.15.4}

In this section we propose a novel analytical setup to derive the network
performance indicators, namely the reliability as probability of
successful packet reception, the delay for successfully received
packets, and the average node energy consumption. We first consider
a single-hop case, and then we generalize the model to
the multi-hop case.


\subsection{Unslotted IEEE~802.15.4 MAC Mechanism}\label{subsec:overview}

According to the IEEE~802.15.4 MAC, each link can be in one of the following states: (i) idle state, when the node is waiting for the next packet to be generated;
(ii) backoff state; (iii) clear channel assessment (CCA) state; (iv) transmission state.

Let the link $l$ be in idle state with probability $b^{(l)}_{0,0,0}$. The three variables given by the number of backoffs $NB$, backoff exponent $BE$, and retransmission attempts $RT$ are initialized: the default initialization is $NB:=0$, $BE:=\emph{macMinBE}$, and $RT:=0$. Note that we use the italic for the MAC variables, as these are the conventional names used in the standard~\cite{ieee802154}. From idle state, the transmitting node wakes up with probability $q_l$, which represents the packet generation probability in each time unit of duration $\emph{aUnitBackoffPeriod}$, and moves to the first backoff state, where the node waits for a random number of complete backoff periods in the range $[0, 2^{BE}-1]$ time units.

When the backoff period counter reaches zero, the node performs the CCA procedure. If the CCA fails due to busy channel, the value of both $NB$ and $BE$ is increased by one. Once $BE$ reaches its maximum value \emph{macMaxBE}, it remains at the same value until it is reset. If $NB$ exceeds its threshold \emph{macMaxCSMABackoffs}, the packet is discarded due to channel access failure. Otherwise the CSMA/CA algorithm generates again a random number of complete backoff periods and repeats the procedure.
The link is in CCA state with probability $\tau_l$, and either moves to the next backoff state if the channel is sensed busy with probability $\alpha_l$, or moves to transmission state with probability ($1-\alpha_l$). The transmitting node experiences a delay of \emph{aTurnaroundTime} to turn around from listening to transmitting mode.

The reception of the corresponding ACK is interpreted as successful packet transmission.
The link moves from the transmission state to idle state with probability ($1-\gamma_{l}$). As an alternative, with probability $\gamma_{l}$, the packet is lost
and the variable $RT$ is increased by one. As long as $RT$ is less than its threshold
\emph{macMaxFrameRetries}, the MAC layer initializes $BE:=\emph{macMinBE}$ and starts again the CSMA/CA mechanism to re-access the channel. Otherwise the packet is discarded as the retry limit is exceeded.

In the following, we denote the MAC parameters by $m_{0}
\triangleq \emph{macMinBE}$, $m_{b} \triangleq \emph{macMaxBE}$, $m
\triangleq \emph{macMaxCSMABackoffs}$, $n \triangleq
\emph{macMaxFrameRetries}$, and $S_b \triangleq   \emph{aUnitBackoffPeriod}$.

\subsection{MAC-Physical Layer Model}\label{sec:model}

In this subsection, the MAC model presented in~\cite{PG_TVT}, which was developed for ideal channel conditions, is substantially modified and extended to include the main features of real channel impairments and interference.

Let us assume that packets are generated by node $i$ according to the Poisson distribution with rate $\lambda_i$. The probability of generation of a new packet after an idle unit time is then $q_l=1 - \exp(-\lambda_i/S_b)$. The effects of a limited buffer size can be included for each link $l$, by considering the probability that the node queue is not empty i) after a packet has been successfully sent $q_{\mathrm{succ},l}\,$, ii) after a packet has been discarded due to channel access failure $q_{\mathrm{cf},l}$ or iii) due to the retry limit $q_{\mathrm{cr},l}$.

We define the packet successful transmission time $L_{s}$
and the packet collision time $L_{c}$ as
\begin{align}
L_{s} &= L + t_{\rm ack} + L_{\rm ack} + IFS \,, \nonumber  \\
L_{c} &= L + t_{\rm m,ack}\,, \label{eq:length}
\end{align}
where $L$ is the total length of a packet including overhead and
payload, $t_{\rm ack}$ is ACK waiting time, $L_{\rm ack}$ is the
length of ACK frame, \emph{IFS} is the inter-frame spacing, and $t_{\rm m,ack}$ is the timeout (waiting for the ACK) in the retransmission algorithm, as detailed in~\cite{ieee802154}.


By using Proposition 4.1 in~\cite{PG_TVT}, the CCA probability $\tau_l$ can be expressed as a function of the packet generation probability $q_l$, the busy channel probability $\alpha_l$, the packet loss probability $\gamma_l$, and the MAC parameters $m_0, m_b, m$, and $n$ as
\begin{align}\label{eq:tau}
\tau_l =\left(\frac{1-\alpha_l^{m+1}}{1-\alpha_l}\right)
\left(\frac{1-\xi_l^{n+1}}{1-\xi_l}\right) b^{(l)}_{0,0,0}\,,
\end{align}
where
\begin{align}\label{eq:b000}
b^{(l)}_{0,0,0}= \left \{
\begin{array} {l}
\left[ \frac{1}{2}\left(
\frac{1-(2\alpha_l)^{m+1}}{1-2\alpha_l}2^{m_{0}}
+\frac{1-\alpha_l^{m+1}}{1-\alpha_l}\right) \frac{1-\xi_l^{n+1}}{1-\xi_l} \right.
    + (L_{s}(1-\gamma_{l}) + L_{c}\gamma_{l})(1-\alpha^{m+1}) \frac{1-\xi_l^{n+1}}{1-\xi_l} \\ \left.
  + \frac{1-q_{cf,l}}{q_l}\frac{\alpha_l^{m+1} (1-\xi_l^{n+1})}{1-\xi_l} +
\frac{1-q_{cr,l}}{q_l} \xi_l^{n+1}
    + \frac{1-q_{\mathrm{succ},l}}{q_l}(1-\gamma_{l})\frac{(1-\alpha_l^{m+1})(1-\xi_l^{n+1})}{1-\xi_l}\right]^{-1}\,,\\
\hfill \mathrm{if} \; m \leq \bar{m} = m_{b}-m_{0}\,,  \\
\left[ \frac{1}{2}\left(
\frac{1-(2\alpha_l)^{\bar{m}+1}}{1-2\alpha_l}2^{m_{0}} +
\frac{1-\alpha_l^{\bar{m}+1}}{1-\alpha_l} + \right.\right.
\left.\left.(2^{m_{b}}+1)\alpha_l^{\bar{m}+1}\frac{1-\alpha_l^{m-\bar{m}}}{1-\alpha_l}
\right)\frac{1-\xi_l^{n+1}}{1-\xi_l}\right.\\
    + (L_{s}(1-\gamma_{l}) + L_{c}\gamma_{l})(1-\alpha_l^{m+1}) \frac{1-\xi_l^{n+1}}{1-\xi_l} + \frac{1-q_{cf,l}}{q_l}\frac{\alpha_l^{m+1} (1-\xi_l^{n+1})}{1-\xi_l}  \\
   \left.+ \frac{1-q_{cr,l}}{q_l}\xi_l^{n+1}
    + \frac{1-q_{\mathrm{succ},l}}{q_l}(1-\gamma_{l})\frac{(1-\alpha_l^{m+1})(1-\xi_l^{n+1})}{1-\xi_l}\right]^{-1}\,,\hfill \mathrm{otherwise,}\\
\end{array}
\right.  \end{align}
and $\xi_l=\gamma_{l}(1-\alpha_l^{m+1})$.

The expressions of the idle state probability in Eq.~\eqref{eq:b000} and the CCA probability in Eq.~\eqref{eq:tau} abstract the behavior of the MAC independently of the underlying physical layer and channel conditions, that we include in the following by deriving novel expressions of the busy channel probability $\alpha_l$ and packet loss probability $\gamma_{l}$.

The busy channel probability can be decomposed as
\begin{align}\label{eq:alpha}
\alpha_{l} = \alpha_{{\rm pkt},l} + \alpha_{{\rm ack},l}\,,
\end{align}
where $\alpha_{{\rm pkt},l}$ is the probability that node $i$
senses the channel and finds it occupied by an ongoing packet
transmission, whereas $\alpha_{{\rm ack},l}$ is the probability of
finding the channel busy due to ACK transmission. Next we derive these probabilities.

The busy channel probability due to packet transmissions evaluated
at node $i$ is the combination of three events:
\begin{enumerate} %
\item at least one other node has accessed the channel within one of the
previous $L$ units of time; %
\item at least one of the nodes that had accessed the
channel found it idle and started a transmission; %
\item the total received power at node $i$ is larger than a
threshold $a$, so that an ongoing transmission is detected by
node $i$.%
\end{enumerate} 

The combination of all busy channel events yields the busy
channel probability that the transmitting node $i$ in link $l$ senses the channel and finds it
occupied by an ongoing packet transmission
\begin{align}\label{eq:alpha_pkt}
\alpha_{{\rm pkt},l} = L\, \bold{\mathcal{H}}_l\left(p^{\mathrm{det}}_{i}\right)\,,
\end{align}
where
\begin{align}
\bold{\mathcal{H}}_l(\chi) =& \sum_{v=1}^{N-1}
\sum_{j=1}^{C_{N-1,v}}\prod_{k=1}^{v} \tau_{k_j} \prod_{h=v+1}^{N-1}
(1-\tau_{h_j})  \sum_{x=1}^{v}
\sum_{n=1}^{C_{v,x}}\prod_{z=1}^{x}(1- \alpha_{z_n})\,
\chi  \prod_{r=x+1}^{v} \alpha_{r_n} \,,
\end{align}
$$
C_{N-1,v}=\binom{N-1}{v} \,,
$$
and 
\begin{align}
p^{\mathrm{det}}_{i} =&\Pr\left[\sum_{z=1}^{x} P_{\mathrm{rx},z_n,i} > a\right]\,
\end{align}
is the detection probability.
The index $v$ accounts for the events of simultaneous accesses to
the channel and the index $j$ enumerates the combinations of
events in which a number $v$ of channel accesses are performed in
the network simultaneously. Given $N$ nodes in the network, the index $k_j$ refers to
the node in the $k$-th position in the $j$-th combination of $v$
out of $N-1$ elements (node $i$ is not included).
 The index $x$
accounts for the events of idle channel, and the index $n$ accounts for the combinations of events in which one or more nodes
among $v$ nodes that access the channel 
find the channel idle simultaneously.

The busy channel probability due to an ACK transmission, recall Eq.~\eqref{eq:alpha}, follows from
a similar derivation. An ACK is sent only after a
successful packet transmission. Therefore,
\begin{align}\label{eq:alpha_ack}
\alpha_{{\rm ack},l} = L_{\rm ack}
\bold{\mathcal{H}}_l\left((1-\gamma_{q_n,w})
p^{\mathrm{det}}_{i}\right)\,,
\end{align}
where $L_{\rm ack}$ is the length of the ACK. The index $w$
denotes the destination node of $q_n$ in the expression of
$\bold{\mathcal{H}}_l$. By summing up Eqs.~\eqref{eq:alpha_pkt} and~\eqref{eq:alpha_ack}, we compute
$\alpha_l$ in Eq.~\eqref{eq:alpha}.


We next derive an expression for the packet loss probability
$\gamma_{l}$, namely the probability that a transmitted packet from node $i$ is not
correctly detected in reception by node $j$. 
A packet transmission is not detected in reception if there is at
least one interfering node that starts the transmission at the same
time and the SINR between the
received power from the intended transmitter and the total interfering
power plus the noise power  level $N_0$ is lower than a threshold $b$
(outage). In the event of no active interferers, which occurs with
probability 1-$\bold{\mathcal{H}}_l(1)$, the packet loss
probability is the probability that the signal-to-noise ratio
(SNR) between the received power and the noise level is lower than
$b$. Hence,
\begin{align}\label{eq:gamma}
\gamma_{l} =(1-\bold{\mathcal{H}}_l(1))\,p^{\mathrm{fad}}_{l} +\bold{\mathcal{H}}_l\left(p^{\mathrm{out}}_{l}\right)+ (2\,L-1)\, \bold{\mathcal{H}}_l\left(\left(1- p^{\mathrm{det}}_{l}\right) p^{\mathrm{out}}_{l} \right)\,,
\end{align}
where $p^{\mathrm{fad}}_{l}$ is the outage probability due to composite channel fading on the useful link (with no
interferers),
\begin{align}
p^{\mathrm{fad}}_{l}= \Pr\left[\frac{P_{\mathrm{rx},l}}{N_0} < b\right]\,,
\end{align}
and $p^{\mathrm{out}}_{l}$ is the outage probability in the presence of interferers (with composite and different channel fading on every link),
\begin{align}
p^{\mathrm{out}}_{l}= \Pr\left[\frac{P_{\mathrm{rx},l}}{\sum_{q=1}^{x}P_{\mathrm{rx},q_n,j} + N_0} < b\right]\,.
\end{align}

The expressions of the carrier sensing probability $\tau_l$ in
Eq.~\eqref{eq:tau}, the busy channel probability $\alpha_l$ in
Eq.~\eqref{eq:alpha}, the collision probability in
Eq.~\eqref{eq:gamma}, for~$l=1,\ldots,N$, form a system of
non-linear equations that can be  solved through
numerical methods~\cite{tsi}.

We next need to derive the detection probability and the outage
probabilities in the devised wireless context. With such a goal in mind, we present some useful
intermediate results in the following section.

\subsection{Model of Aggregate Multi-path Shadowed Signals} \label{subsec:model-fading}

In this section, we consider the problem of computing the sum of multi-path shadowed signals that appear in the detection probability and in the outage
probability.  The analysis follows the approach developed in~\cite{pratesiTWC} and \cite{fischioneTCOMM} for cellular systems, adapting the model to the characteristics of CSMA/CA systems.  

Consider the transmitting node $i$ performing a CCA and let us focus our attention on
the detection probability in transmission $\Pr\left[\sum_{n=1}^{x}
P_{\mathrm{rx},n,i} > a\right]$, where $x$ is the current number
of active nodes in transmission.
By recalling the power channel model in Eq.~\eqref{eq:prx},
let us define the random variable
$Y_i=\ln\left(\sum_{n=1}^x A_{i,n} \exp(y_n)\right)$,
with $A_{i,n}= c_0 P_{\mathrm{tx},n}\,f_n/r_{n,i}^k$, and
$y_n\sim\mathcal{N}(0,\sigma_n^2)$. Since a closed form expression of
the probability distribution function of $Y_i$ does not exist, we
resort to a useful approximation instead. In order to characterize
$Y_i$, we apply the Moment Matching Approximation (MMA) method, which approximates the statistics
of linear combination of lognormal components with a lognormal
random variable,  such that $Y_i\sim\mathcal{N}(\eta_{Y_{i}},\sigma_{Y_i}^2)$.
According to the MMA method, $\eta_{Y_i}$ and $\sigma_{Y_i}$ can
be obtained by matching the first two moments of $\exp(Y_i)$ with
the first two moments of $\sum_{n=1}^{x} A_{i,n} \exp(y_n)$, i.e.,
\begin{align}\label{eq:mma1}
M_1  \triangleq \exp\left(-\eta_{Y_i}+\frac{1}{2}\sigma_{Y_i}\right) =\sum_{n=1}^{x} E\{A_{i,n}\} \exp\left(\eta_{y_n}+\frac{1}{2}\sigma_{y_n}\right)\,,\end{align}
\begin{align}\label{eq:mma2} M_2 \triangleq  \exp \left(- 2 \eta_{Y_i}+
2\sigma_{Y_i}\right)\hspace{-2pt}=\hspace{-3pt}\sum_{m=1}^{x}\sum_{n=1}^x E\{A_{i,m} A_{i,n}\}
\exp \hspace{-3pt} \left(\eta_{y_m}\hspace{-5pt}+\hspace{-3pt}\eta_{y_n} \hspace{-5pt}+ \hspace{-5pt}\left(\frac{\sigma^2_{y_m}}{2}+\frac{\sigma^2_{y_n}}{2}+\rho_{y_m,y_n}\sigma_{y_m}\sigma_{y_n}\right)\hspace{-3pt}\right)\hspace{-3pt}.
\end{align}
Solving Eqs.~\eqref{eq:mma1}, and~\eqref{eq:mma2} for $\eta_{Y_i}$
and $\sigma_{Y_i}$ yields
$\eta_{Y_i}=0.5\ln(M_2)-2\ln(M_1)$, and $ \sigma^2_{Y_i}=\ln(M_2)-2\ln(M_1)$.

It follows that
\begin{align} \label{eq:outrx}
p^{\mathrm{det}}_{i}=\Pr\left[\sum_{n=1}^{x} P_{\mathrm{rx},n,i} > a\right] = \Pr\left[\exp(Y_i) > a\right] \approx Q
\left( \frac{\ln(a)-\eta_{Y_{i}}}{\sigma_{Y_{i}}} \right)\,,
\end{align}
where $Q(z)=\frac{1}{\sqrt{2 \pi}} \int_{z}^{\infty} \exp\left(-\frac{\nu^2}{2}\right) d\nu.$

Similar derivations follow for the outage probability in
reception
\begin{align*}
&\Pr \hspace{-3pt} \left[\frac{P_{\mathrm{rx},i,j}}{\sum_{n=1}^{x}P_{\mathrm{rx},n,j} + N_0}\hspace{-2pt} < b\right] \hspace{-3pt}=\hspace{-3pt} \Pr \hspace{-3pt} \left[ f_i \left(\sum_{n=1}^{x}\frac{P_{\mathrm{tx},n}r_{i,j}^k}{P_{\mathrm{tx},i} r_{n,j}^k} f_n \exp(y_n-y_i)+\frac{N_0 r_{i,j}^k}{P_{\mathrm{tx},i}} f_n \exp(-y_i) \right)^{-1}\hspace{-9pt}
< b\right]\hspace{-3pt}.
\end{align*}
Let us now define the random variable
$\tilde{Y}_{i,j}=-\ln\left(\sum_{n=1}^{x+1} B_{i,j,n}
\exp(\tilde{y}_n)\right)\,,
$ where 
$$
B_{i,j,n}=\left\{\begin{array}{cl}\frac{P_{\mathrm{tx},n}r_{i,j}^k}{P_{\mathrm{tx},i} r_{n,j}^k} f_n & \mathrm{for}\quad n=1,...,x\\
\frac{N_0 r_{i,j}^k}{P_{\mathrm{tx},i}} f_n & \mathrm{for}\quad n=x+1 \end{array}
 \right.  \,, \qquad
\tilde{y}=\left\{\begin{array}{cl} y_n - y_i & \mathrm{for}\quad n=1,...,x\\
- y_i  & \mathrm{for}\quad n=x+1 \end{array}
 \right.\,.
$$

According to the MMA method, we approximate $\tilde{Y}_i\sim\mathcal{N}(\eta_{Y_{i}},\sigma_{Y_i}^2)$, where $\eta_{\tilde{Y}_{i,j}}$ and
$\sigma_{\tilde{Y}_{i,j}}$ can be obtained by matching the first two
moments of $\exp(\tilde{Y}_i)$ with the first two moments of
$\sum_{n=1}^{N} B_{i,j,n} \exp(\tilde{y}_n)$, i.e.,
\begin{align*}
\tilde{M}_1 \triangleq & \exp\left(-\eta_{\tilde{Y}_{i,j}}+\frac{1}{2}\sigma_{\tilde{Y}_{i,j}}\right)\hspace{-3pt}=\hspace{-3pt}\sum_{n=1}^{x+1} E\{B_{i,j,n}\} \exp\left(\eta_{\tilde{y}_n}+\frac{1}{2}\sigma_{\tilde{y}_n}\hspace{-3pt}\right),\end{align*}
\begin{align*} \tilde{M}_2 \triangleq &\exp(- 2 \eta_{\tilde{Y}_{i,j}}\hspace{-0.15cm}+
2\sigma_{\tilde{Y}_{i,j}})\hspace{-0.1cm}=\hspace{-0.15cm} \sum_{m=1}^{x+1}\sum_{n=1}^{x+1}\hspace{-3pt} E\{B_{i,j,m} B_{i,j,n}\}
\hspace{-3pt}\exp\hspace{-3pt}\left(\eta_{\tilde{y}_m} \hspace{-0.1cm} +\hspace{-0.1cm}\eta_{\tilde{y}_n} \hspace{-0.1cm}+\hspace{-0.1cm}
\left(\frac{\sigma^2_{\tilde{y}_m}}{2}\hspace{-0.1cm}+\hspace{-0.1cm}\frac{\sigma^2_{\tilde{y}_n}}{2}\hspace{-0.1cm}+\hspace{-0.1cm}\rho_{\tilde{y}_m,\tilde{y}_n}\sigma_{\tilde{y}_m}\sigma_{\tilde{y}_n}\right)\hspace{-3pt}\right)\hspace{-3pt},
\end{align*}
which yields $\eta_{\tilde{Y}_{i,j}}=0.5\ln(\tilde{M}_2)-2\ln(\tilde{M}_1)$,
$\sigma^2_{\tilde{Y}_{i,j}}=\ln(\tilde{M}_2)-2\ln(\tilde{M}_1)$. Therefore,
\begin{align}  p^{\mathrm{out}}_{i,j}
&=  \Pr\left[ f_i \exp(\tilde{Y}_{i,j}) < b\right]  = \int_0^b \int_0^{\infty}p_f(z | w) p_{\exp(\tilde{Y}_{i,j})}(w) dw\, dz \nonumber\\
&= \int_0^b \int_0^{\infty}p_f(z | w) \frac{1}{\sqrt{2 \pi} \sigma_{\tilde{Y}_{i,j}} w} \exp\left(-\frac{(\ln(w)-\eta_{\tilde{Y}_{i}})^2}{2 \sigma^2_{\tilde{Y}_{i}}} \right)  dw \,dz\,.
\end{align}

The analysis above holds for a generic weighted composition of lognormal fading components. In the case of lognormal channel model, where only shadow fading components are considered, (i.e., $f_i=1$), the outage probability becomes
\begin{align} \label{eq:poutrxb} p^{\mathrm{out,L}}_{i,j} =
\Pr\left[\exp(\tilde{Y}_{i,j}) < b\right]\approx 1-Q \left(
\frac{\ln(b)-\eta_{\tilde{Y}_{i,j}}}{\sigma_{\tilde{Y}_{i,j}}}
\right)\,.
\end{align}
For a Nakagami-lognormal channel, the outage probability becomes
\begin{align*} p^{\mathrm{out,NL}}_{i,j}
&= \int_0^b \int_0^{\infty}  \kappa^\kappa \frac{(z w)^{\kappa-1}}{\Gamma(\kappa)} \exp(-\kappa z w)  \frac{1}{\sqrt{2 \pi} \sigma_{\tilde{Y}_{i,j}} w} \exp\left(-\frac{(\ln(w)-\eta_{\tilde{Y}_{i}})^2}{2 \sigma^2_{\tilde{Y}_{i}}} \right)  dw \,dz\ \nonumber \\ &= \int_0^{\infty}  \frac{1}{\sqrt{2 \pi} \sigma_{\tilde{Y}_{i,j}} w} \exp\left(-\frac{(\ln(w)-\eta_{\tilde{Y}_{i}})^2}{2 \sigma^2_{\tilde{Y}_{i}}} \right)    \int_0^b \kappa^\kappa \frac{(z w)^{\kappa-1}}{\Gamma(\kappa)} \exp(-\kappa z w) dz  \,dw\,.
\end{align*}
For integer values of $m$, the integration in $z$ yields
\begin{align*}  p^{\mathrm{out,NL}}_{i,j}
= 1 - \int_0^{\infty}  \frac{1}{\sqrt{2 \pi} \sigma_{\tilde{Y}_{i,j}} w} \exp\left(-\frac{(\ln(w)-\eta_{\tilde{Y}_{i}})^2}{2 \sigma^2_{\tilde{Y}_{i}}} \right) \sum_{i=0}^{\kappa-1}  \frac{(\kappa \, b\, w)^i}{\Gamma(i+1)}\exp(-\kappa\, b\, w) dw\,.
\end{align*}

The mean and standard deviation of $Y_i$ and $\tilde{Y}_{i,j}$ can be obtained by inserting the moments of $f_i$ in the moments of $A_{i,n}$ and $B_{i,j,n}$.
For Gamma distributed components $f_i$, we obtain $ E\{f_i\} = 1$ and $E\{f^2_i\}=(\kappa+1)/\kappa$.

We remark here that the evaluation of $p^{\mathrm{det}}_{i}$ and $p^{\mathrm{out}}_{i,j}$ can be carried out off-line with respect to the solution of the system of nonlinear equations that need to be solved when deriving $\tau_l$, $\alpha_l$ and $\gamma_{l}$. Therefore, the proposed model can be implemented with only a slight increase of complexity with respect to the analytical model of the IEEE~802.15.4 MAC mechanism presented in~\cite{PG_TVT}, but the online computation time is not affected significantly.

\subsection{Extended Model for Multi-hop Networks}\label{sec:multihop}
Here we extend the analytical model to a general
network in which information is forwarded through a multi-hop
communication towards a sink node.

The model equations derived in Section~\ref{sec:model} are solved for each link of the network, by considering
that the probability $q_l$ of having a packet to transmit in each time unit 
does not depend only on the generated traffic $\lambda_i$ from the transmitting node $i$,
but also on the traffic to forward from children nodes according to the routing policy.

The effect of routing can be described by the routing matrix
$\textbf{M}$, such that
$M_{i,j}=1$ if node $j$ is the destination of node $i$, and
$M_{i,j}=0$ otherwise. We assume that the routing matrix is built
such that no cycles exists. We define the traffic distribution
matrix $\textbf{T}$ by scaling $\textbf{M}$ by the probability of
successful reception in each link as only successfully received
packets are forwarded, i.e., $T_{i,j}= M_{i,j} R_{l}$,
where the reliability $R_{l}$ is derived next in Section~\ref{subsec:reliab}.
The vector of traffic generation probabilities $Q
$ is then given in~\cite{PG_TVT} by
\begin{align}\label{eq:q}
Q = \lambda\, [\textbf{I} - {\textbf{T}}]^{-1}  \,.
\end{align}
where \textbf{I}$\in \mathds{R}^{(N+1) \times (N+1)}$ is the
identity matrix.  Eq.~\eqref{eq:q} gives the relation between MAC and routing through the idle packet
generation probability $q_l$.
To include the effects of fading channels in the multi-hop network model, we
couple Eq.~\eqref{eq:q} with the expressions for $\tau_l$ and
$\alpha_l$, as obtained by
Eqs.~\eqref{eq:tau}, and~\eqref{eq:alpha}.
Moreover, to complete the model, we need to derive the expression of the reliability $R_l$,
as we illustrate in the following section.

\section{Performance Indicators}\label{sec:reliab}

In this section, we investigate three major indicators to
analyze the performance of the IEEE~802.15.4 MAC over fading
channels. These indicators will also be used to validate
the analytical model we derived in the previous section, by comparing results obtained from the (approximate) model with those obtained by extensive simulation campaigns. The first one is the reliability, evaluated as
successful packet reception rate. Then we consider the delay for
the successfully received packets as the time interval from the
instant the packet is ready to be transmitted, until an ACK for
such a packet is received. Eventually, we consider the energy
consumption of network nodes.

\subsection{Reliability}\label{subsec:reliab}

For each node of the network, the reliability is based on the
probability that packets are discarded at MAC layer. In unslotted
CSMA/CA, packets are discarded due to either (i) channel access failure or (ii) retry limits. A
channel access failure happens when a packet fails to obtain clear
channel within $m+1$ backoff stages in the current transmission attempt. Furthermore, a packet is
discarded if the transmission fails due to repeated packet losses
after $n+1$ attempts. 
According to the IEEE 802.15.4 MAC mechanism described in Section~\ref{subsec:overview}, the probability that the packet is discarded due to channel access
failure can be expressed as
\begin{align*}
p_{{\rm cf},l} & = \alpha_l^{m+1} \sum_{j=0}^{n} (\gamma_{l} (1-\alpha_l^{m+1}))^j \,,
\end{align*}
and the probability of a packet being discarded due to retry limits is
\begin{align*}
p_{{\rm cr},l} & = (\gamma_{l}(1-\alpha_l^{m+1}))^{n+1}\,.
\end{align*}
Therefore, the reliability can be expressed as
\begin{align} \label{eq:reliability}
R_{l} & = 1 - p_{{\rm cf},l} - p_{{\rm cr},l} = 1- \alpha_l^{m+1} \frac{(1-(\gamma_{l}(1-\alpha_l^{m+1}))^{n+1})}{1-\gamma_{l}(1-\alpha_l^{m+1})}- (\gamma_{l}(1-\alpha_l^{m+1}))^{n+1}\,.
\end{align}
It is worthwhile mentioning that the last expressions embed the link between the reliability at the MAC level and the statistical description of wireless channel environment through Eq.~\eqref{eq:gamma} and the analysis of Section~\ref{subsec:model-fading}.

\subsection{Delay}

We define the delay $D_{l}$ for successfully delivered packets in the link $l$. If a packet is discarded due to either the limited
number of backoff stages $m$ or the finite retry limit $n$, its
delay is not included into the average delay.

Let $D_{l,h}$ be the delay for the transmitting node that sends a packet
successfully at the $h$-th attempt. The expected value of the delay is
\begin{align}\label{eq:average_delay}
\mathbb{E}[D_{l}] = \mbox{$\sum\limits_{h=0}^{n}$}
\Pr[\mathscr{C}_h|\mathscr{C}] \, \mathbb{E}[D_{l,h}]\,,
\end{align}
where the event $\mathscr{C}_h$ denotes the
occurrence of a successful packet transmission at time $h+1$ given
$h$ previous unsuccessful transmissions, whereas the event
$\mathscr{C}$ indicates a successful packet
transmission within $n$ attempts. Therefore, we can derive
\begin{align}\label{eq:Pr_A}
\Pr[\mathscr{C}_h|\mathscr{C}] &= \frac{\gamma_{l}^{j}(1-\alpha_l^{m+1})^{j}}{\sum_{k=0}^{n}\left(\gamma_{l}(1-\alpha_l^{m+1})\right)^{k}}=\frac{\left(1-\gamma_{l}(1-\alpha_l^{m+1})\right)\gamma_{l}^{j}(1-\alpha_l^{m+1})^{j}
}{1-\left(\gamma_{l}(1-\alpha_l^{m+1})\right)^{n+1}}\,.
\end{align}
We recall that $\gamma_{l}$ is the packet loss probability, which
we derived in Eq.~\eqref{eq:gamma} together with
Eqs.~\eqref{eq:outrx} and~\eqref{eq:poutrxb}, and
$1-\alpha_l^{m+1}$ is the probability of successful channel access
within the maximum number of $m$ backoff stages, where
$\alpha_l^{m+1}$ follows from Eq.~\eqref{eq:alpha}.

The average delay at the $h$-th attempt is
\begin{align}
\mathbb{E}[D_{l,h}] = L_{s} + h\,L_{c} +
\sum_{l=0}^{h} \mathbb{E}[T_{l}] \,,\label{eq:d_l_j}
\end{align}
where $T_{l}$ is the backoff stage delay, whereas $L_{s}$ and $L_{c}$ are the
time periods in number of time units for successful packet
transmission and collided packet transmission computed in Eq.~\eqref{eq:length}.


Since the backoff time in each stage $k$ is uniformly
distributed in $[0,W_{k}-1]$, where $W_k = 2^{BE}$, the expected total backoff delay is
\begin{align}
\mathbb{E}[T_{l}] =& T_{sc} + \mbox{$\sum\limits_{r=0}^{m}$}
\Pr[\mathscr{D}_r|\mathscr{D}] \left(r\,
T_{sc}+\mbox{$\sum\limits_{k=0}^{r}$} \frac{W_{k}-1}{2}\, S_{b}
\right) \,,\label{eq:e_th}
\end{align}
where $T_{sc}$ is the sensing time in the unslotted mechanism.
The event $\mathscr{D}_r$ denotes the occurrence of a busy channel
for $r$ consecutive times, and then an idle channel at the $(r+1)$th
time. By considering all the possibilities of busy channel during
two CCAs, the probability of $\mathscr{D}_r$ is conditioned on the
successful sensing event within $m$ attempts $\mathscr{D}$, given
that the node senses an idle channel in CCA. It follows that
\begin{align}\label{eq:Pr_B}
\Pr[\mathscr{D}_r|\mathscr{D}] =
\frac{\alpha_l^r}{\sum_{k=0}^{m}\alpha_l^k} = \frac{\alpha_l^r
(1-\alpha_l)}{1-\alpha_l^{m+1}}\,.
\end{align}

By applying Eqs.~\eqref{eq:Pr_A}~--~\eqref{eq:Pr_B} in
Eq.~\eqref{eq:average_delay}, the average delay for successfully
received packets is computed. Note that the delay is experienced at the MAC level and is hereby linked to the fading channel through the dependency on $\alpha_l$ and $\gamma_{l}$ evaluated in the previous section.


\subsection{Energy Consumption}

Here we derive the expression of the energy consumption of the transmitting node of link $l$ as
the sum of the contribution in backoff, carrier sense,
transmission, reception, idle-queue, and relay states:
\begin{align} \label{eq:energy}
E_{\mathrm{tot},l}=E_{b,l}+E_{s,l}+E_{t,l}+E_{r,l}+E_{q,l} + E_{x,l}\,.
\end{align}
In the following, each component of this expression is derived according to the
state probabilities in Section~\ref{subsec:overview}. The energy
consumption during backoff is
\begin{align*}
E_{b,l}&= P_{\mathrm{id}} \frac{\tau_l}{2}\left(\frac{(1-(2\alpha_l)^{m+1}) (1-\alpha_l)}{(1-2\alpha_l)(1-\alpha_l^{m+1})}2^{m_{0}} + 1\right)\,,
\end{align*}
where $P_{\mathrm{id}}$ is the average power consumption in
idle-listening state, as we assume that the radio is set in
idle-listening state during the backoff stages. The energy
consumption for carrier sensing is $E_{s,l}= P_{\mathrm{sc}}
\tau_l$, where $P_{\mathrm{sc}}$ is the average node power
consumption in carrier sensing state. The energy consumption
during the transmission stage, including ACK reception, is
\begin{align*}
E_{t,l}=& (1-\alpha_l) \tau_l (P_t L + P_{\mathrm{id}} + (P_r (1-\gamma_{l}) + P_{\mathrm{id}} \gamma_{l}) L_{\mathrm{ack}})\,,
\end{align*}
where $P_{\mathrm{t}}$ and $P_{\mathrm{r}}$ are the average node
power consumption in transmission and reception respectively, and
we assume $t_{\rm m,ack} = L_{\rm ack}+1$ in backoff time units $S_b$.
In the single-hop case, we assume that the node is in sleeping state with
negligible energy consumption during inactivity periods without
packet generation. Hence, the energy consumption during the idle-queue state is given by
$E_{t,l}= P_{\mathrm{s}} \,b^{(l)}_{0,0,0}$,
where $P_{\mathrm{s}}$ is the average node power consumption in sleeping mode,
and $b^{(l)}_{0,0,0}$ is the stationary probability of the idle-queue state as derived in Eq.~\eqref{eq:b000}.

In the multi-hop case, relay
nodes are in idle-listening state also during the inactivity
period (because of the duty cycle policy), and an extra cost for
receiving packets and sending ACKs has to be accounted for.
This is included in the energy consumption $E_{x,i}$ due to the packets and ACKs of relay nodes
based on the routing matrix $\textbf{M}$,
\begin{align*}
E_{x,i}=& \sum_{n=1}^{N} M_{n,i} (1-\gamma_{n,i})(1-\alpha_n) \tau_n (P_t L + P_{\mathrm{id}} + (P_r (1-\gamma_{n,i})+ P_{\mathrm{id}} \gamma_{n,i}) L_{\mathrm{ack}}) \,.
\end{align*}

We validate and show the use of these analytical results in the next section.

\section{Performance Evaluations}\label{sec:simulations}

In this section, we present numerical results for the new model for various settings, network topologies, and operations. We report extensive Monte Carlo simulations  to validate the accuracy of the approximations that we have introduced in the model. 
As discussed in~\cite{Iver09}, the capture threshold model used in the network simulator ns2~\cite{ns2} gives unsatisfactory performance when multiple access interference is considered. Therefore, we implemented the IEEE 802.15.4 MAC mechanism in Matlab. The fading channel conditions are reproduced by generating independent random variables in each link and for each generated packet, and the SINR accounts for the cumulative interference power. In the simulations, we consider that the coherence time of the shadow fading is longer than the packet transmission time, which is in the order of milliseconds, but shorter than the packet generation period, which is in the order of seconds. This is typically true for an IEEE 802.15.4 environment~\cite{ieee802154}.

The setting of the MAC and physical layer parameters is based on the default specifications of the IEEE~802.15.4~\cite{ieee802154}. 
We perform simulations both for single-hop and multi-hop topologies. As a benchmark, we consider the IEEE 802.15.4 MAC model in~\cite{PG_TVT}.
Such a model represent the state of the art for unslotted IEEE 802.15.4 single-hop and multi-hop networks with heterogeneous traffic and hidden terminals. 

\subsection{Single-hop Topologies}

\begin{figure}[t] \centering
\includegraphics[width=0.59\textwidth]{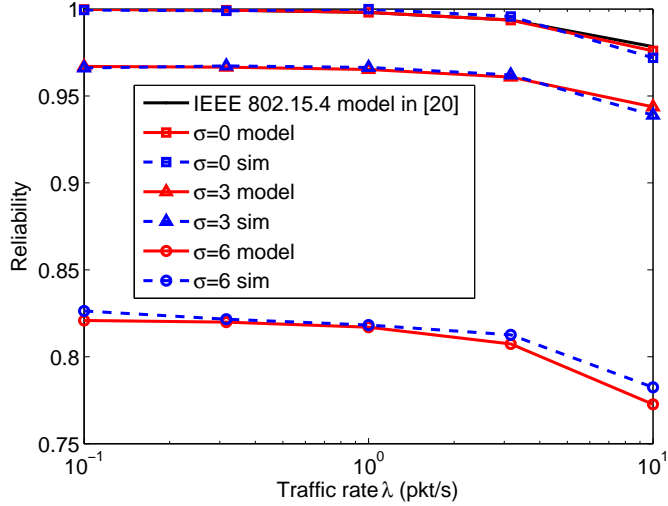}
\caption{Reliability vs. traffic rate $\lambda$ for the star network in Fig.~\ref{fig:topology}a) with $N=7$ nodes, $r=1$~m, $a=-76$~dBm, $b=6$~dB.}\label{fig:rel_lambda}\end{figure}

\begin{figure}[t] \centering
\includegraphics[width=0.59\textwidth]{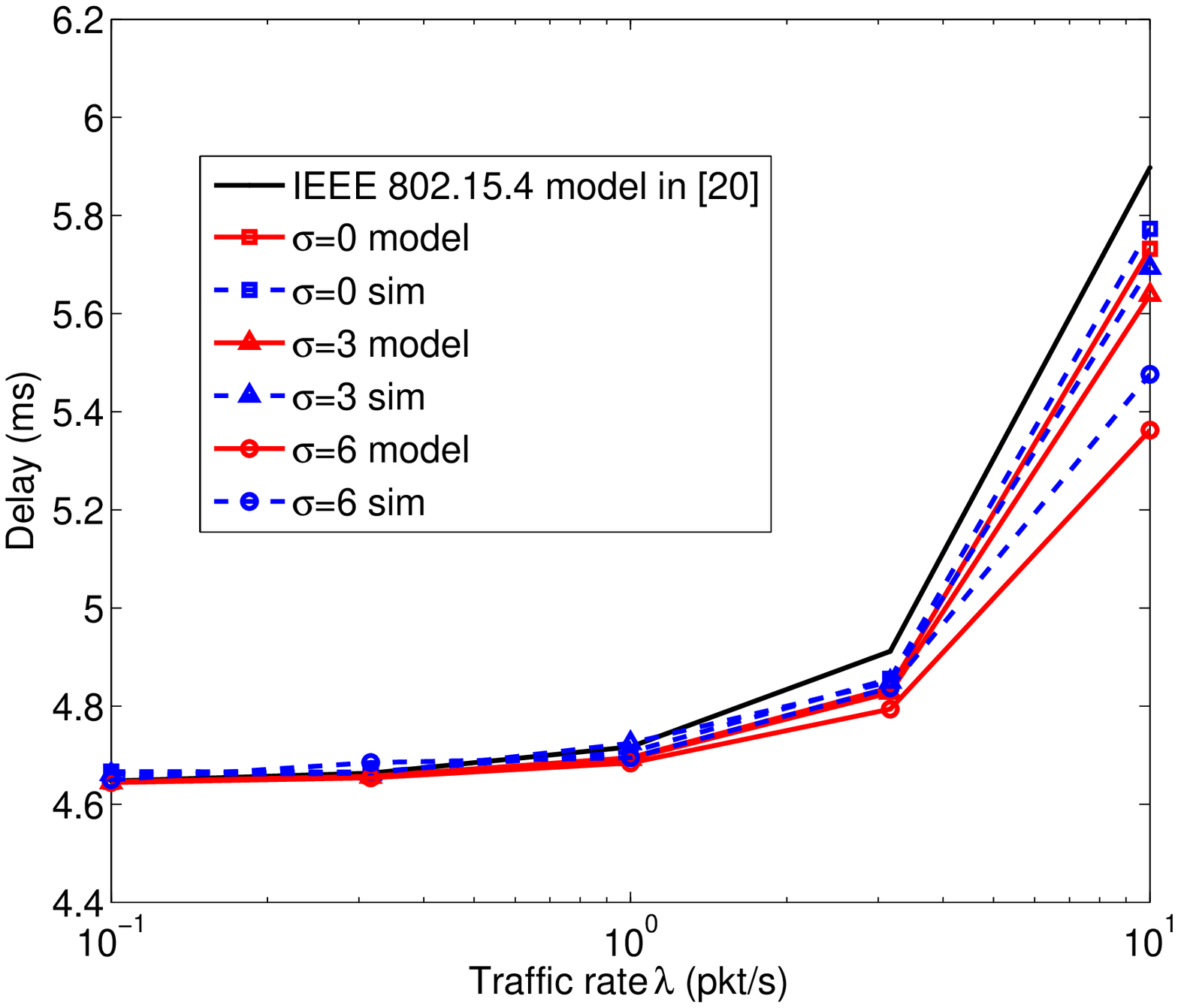}
\caption{Delay vs. traffic rate $\lambda$ for the star network in Fig.~\ref{fig:topology}a) with $N=7$ nodes, $r=1$~m, $a=-76$~dBm, $b=6$~dB.}\label{fig:del_lambda}\end{figure}

\begin{figure}[t] \centering
\includegraphics[width=0.59\textwidth]{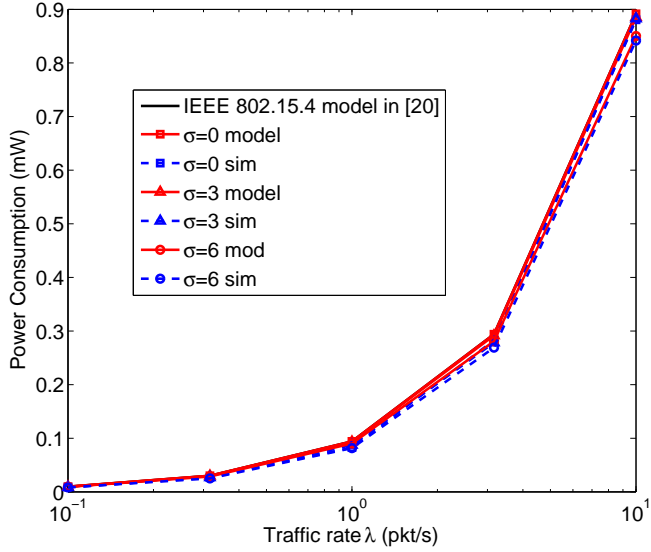}
\caption{Power consumption vs. traffic rate $\lambda$ for the star network in Fig.~\ref{fig:topology}a) with $N=7$ nodes, $r=1$~m, $a=-76$~dBm, $b=6$~dB.}\label{fig:ene_lambda}\end{figure}

\begin{figure}[t] \centering
\includegraphics[width=0.59\textwidth]{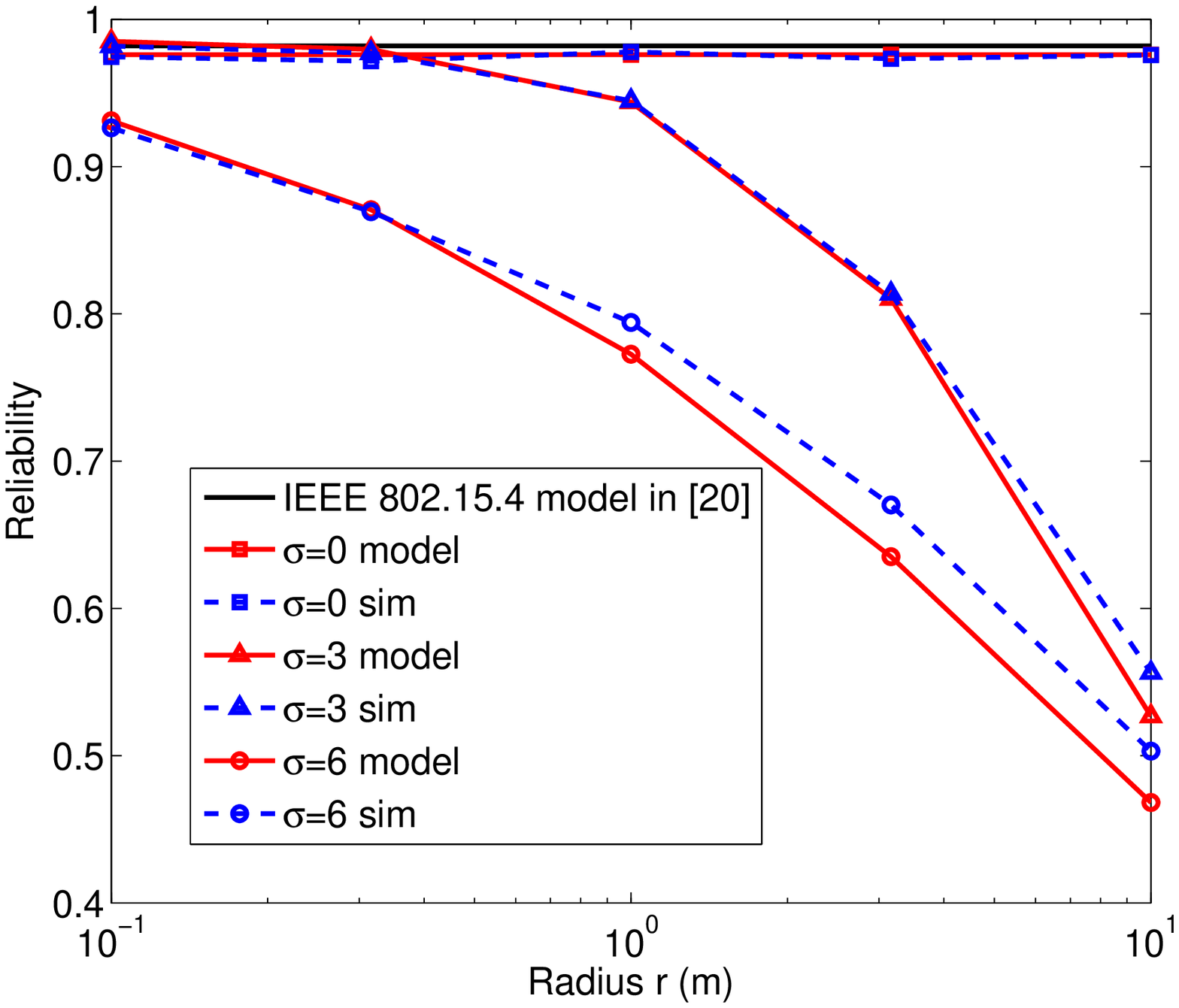}
\caption{Reliability vs. radius $r$ for the star network in Fig.~\ref{fig:topology}a) with $N=7$ nodes, $\lambda=10$~pkt/s, $a=-76$~dBm, $b=6$~dB.}\label{fig:rel_distance}\end{figure}

\begin{figure}[t] \centering
\includegraphics[width=0.59\textwidth]{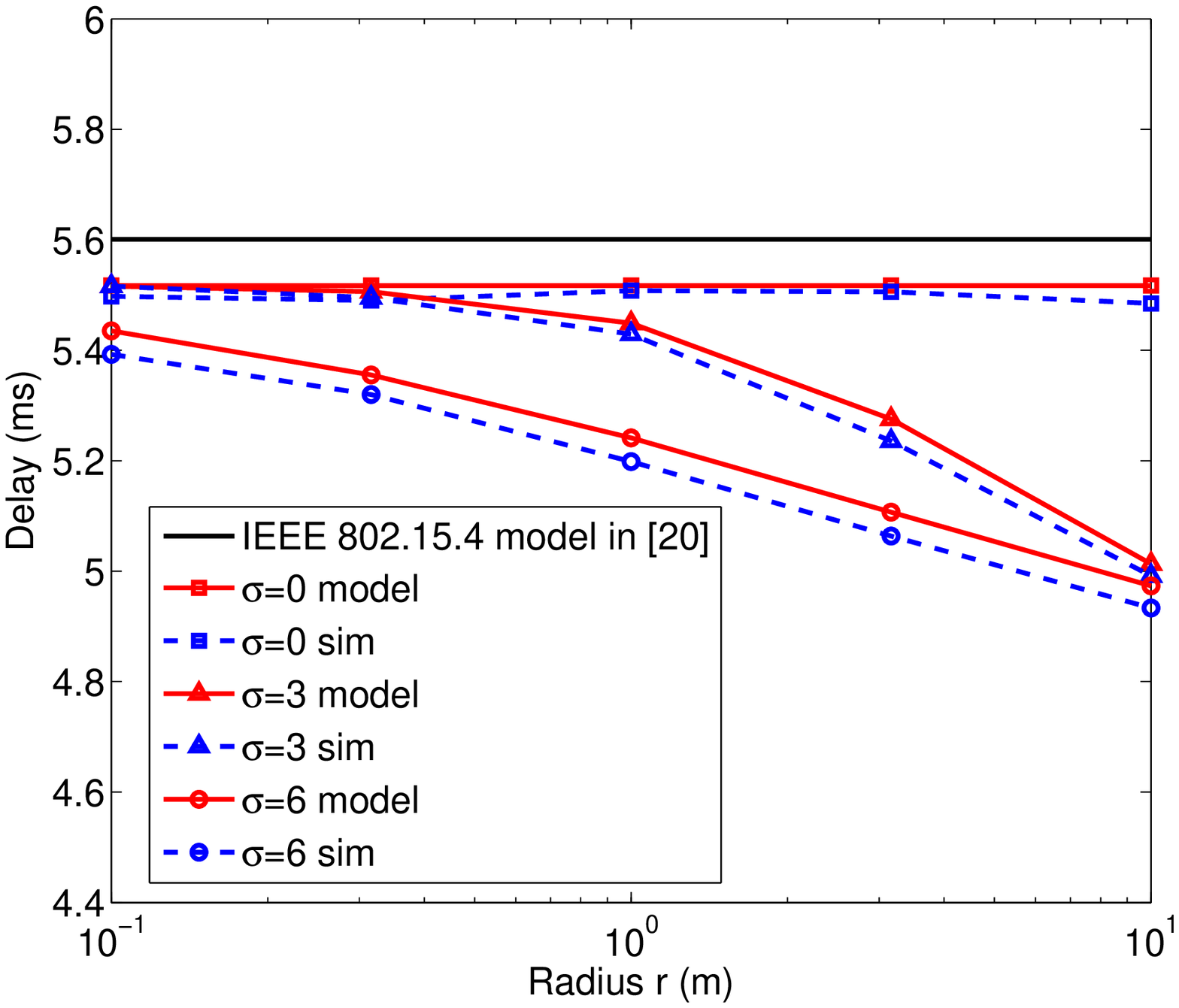}
\caption{Delay vs. radius $r$ for the star network in Fig.~\ref{fig:topology}a)  with $N=7$ nodes, $\lambda=10$~pkt/s, $a=-76$~dBm, $b=6$~dB.}\label{fig:del_distance}\end{figure}

\begin{figure}[t] \centering
\includegraphics[width=0.59\textwidth]{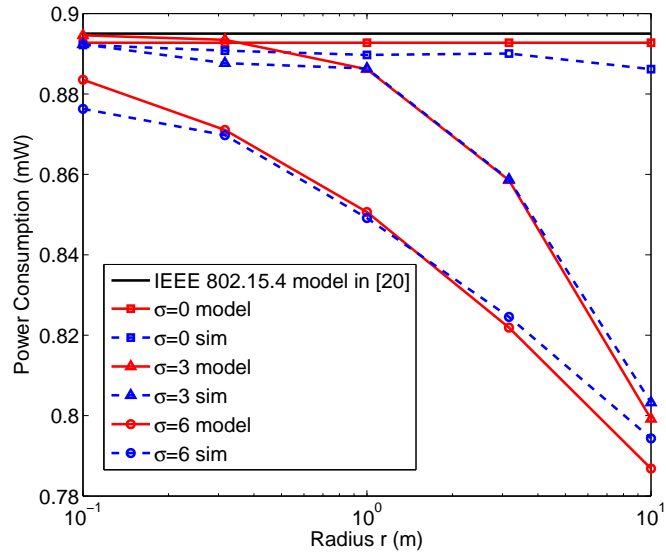}
\caption{Power consumption vs. radius $r$ for the star network in Fig.~\ref{fig:topology}a)  with $N=7$ nodes, $\lambda=10$~pkt/s, $a=-76$~dBm, $b=6$~dB.}\label{fig:ene_distance}\end{figure}

\begin{figure}[t] \centering
\includegraphics[width=0.59\textwidth]{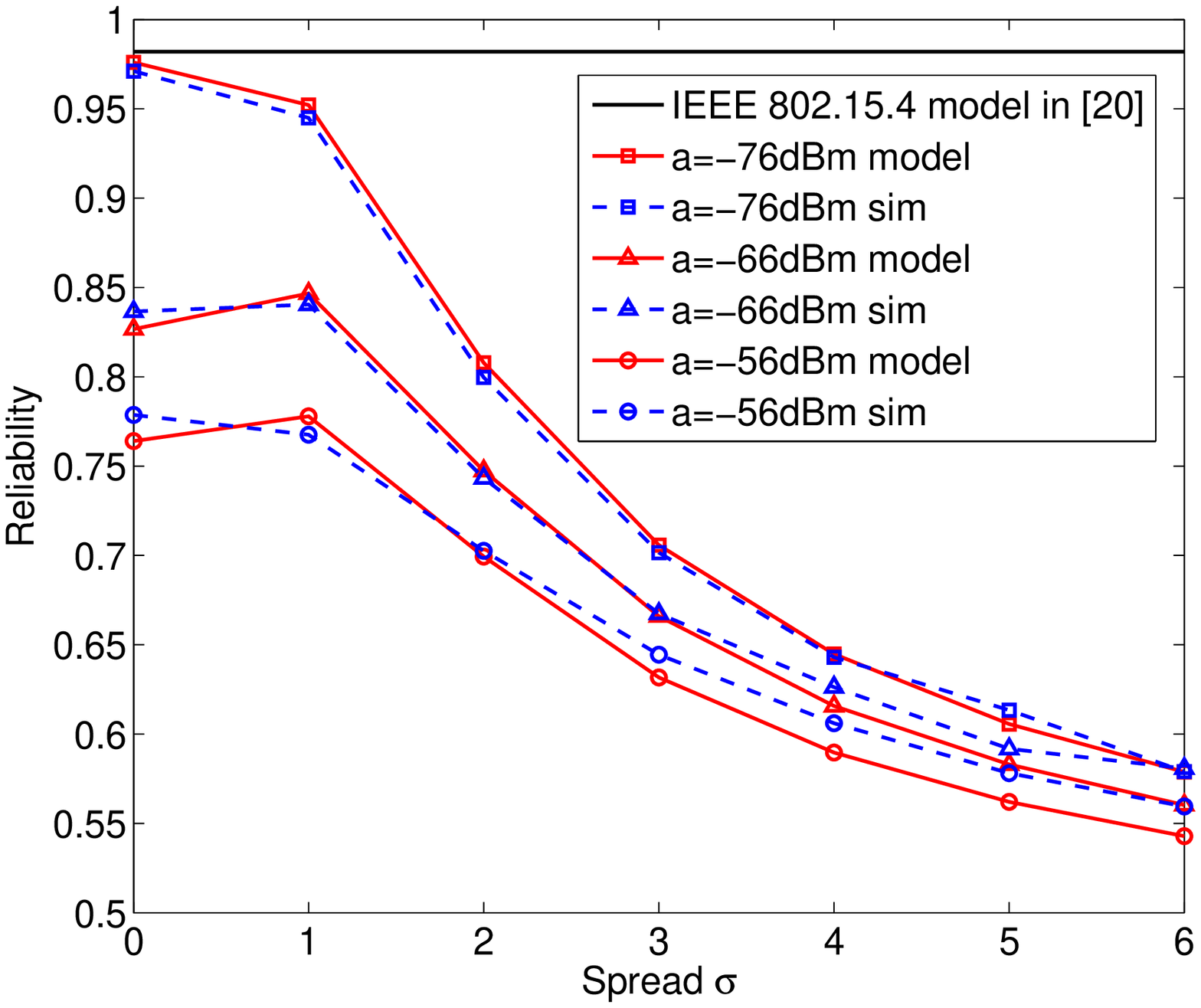}
\caption{Reliability vs. $\sigma$ for the star network in Fig.~\ref{fig:topology}a)  with $N=7$ nodes, $r=5$~m, $\lambda=10$ pkt/s, $b=6$~dB.}\label{fig:rel_sigma_a}\end{figure}

\begin{figure}[t] \centering
\includegraphics[width=0.59\textwidth]{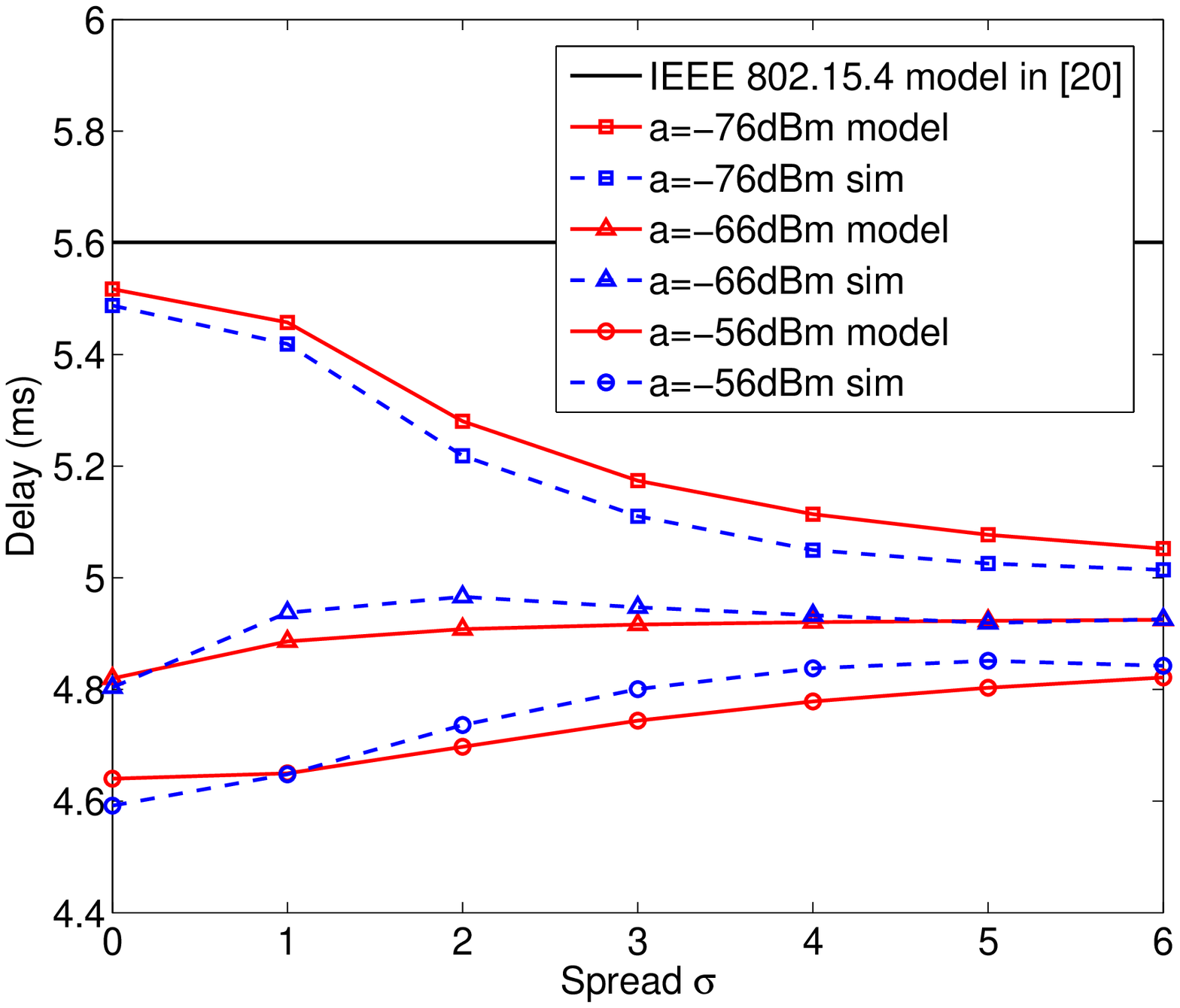}
\caption{Delay vs. $\sigma$ for the star network in Fig.~\ref{fig:topology}a)  with $N=7$ nodes, $r=5$~m, $\lambda=10$ pkt/s, $b=6$~dB.}\label{fig:del_sigma_a}\end{figure}

\begin{figure}[t] \centering
\includegraphics[width=0.59\textwidth]{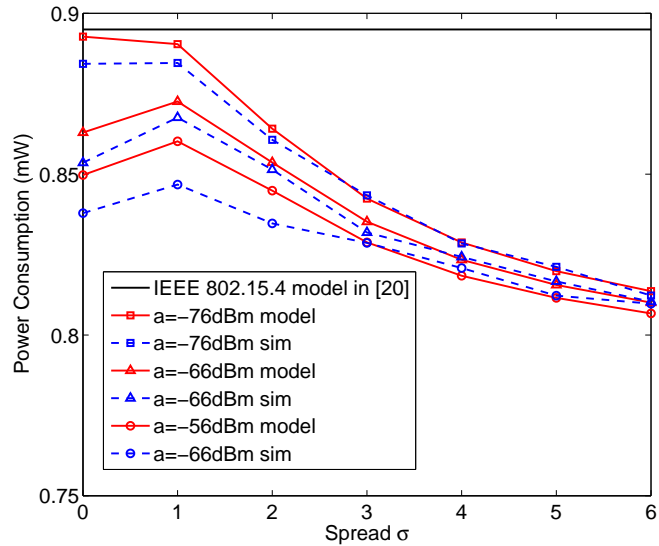}
\caption{Power consumption vs. $\sigma$ for the star network in Fig.~\ref{fig:topology}a)  with $N=7$ nodes, $r=1$~m, $\lambda=10$ pkt/s, $b=6$~dB.}\label{fig:ene_sigma_a}\end{figure}

\begin{figure}[t] \centering
\includegraphics[width=0.59\textwidth]{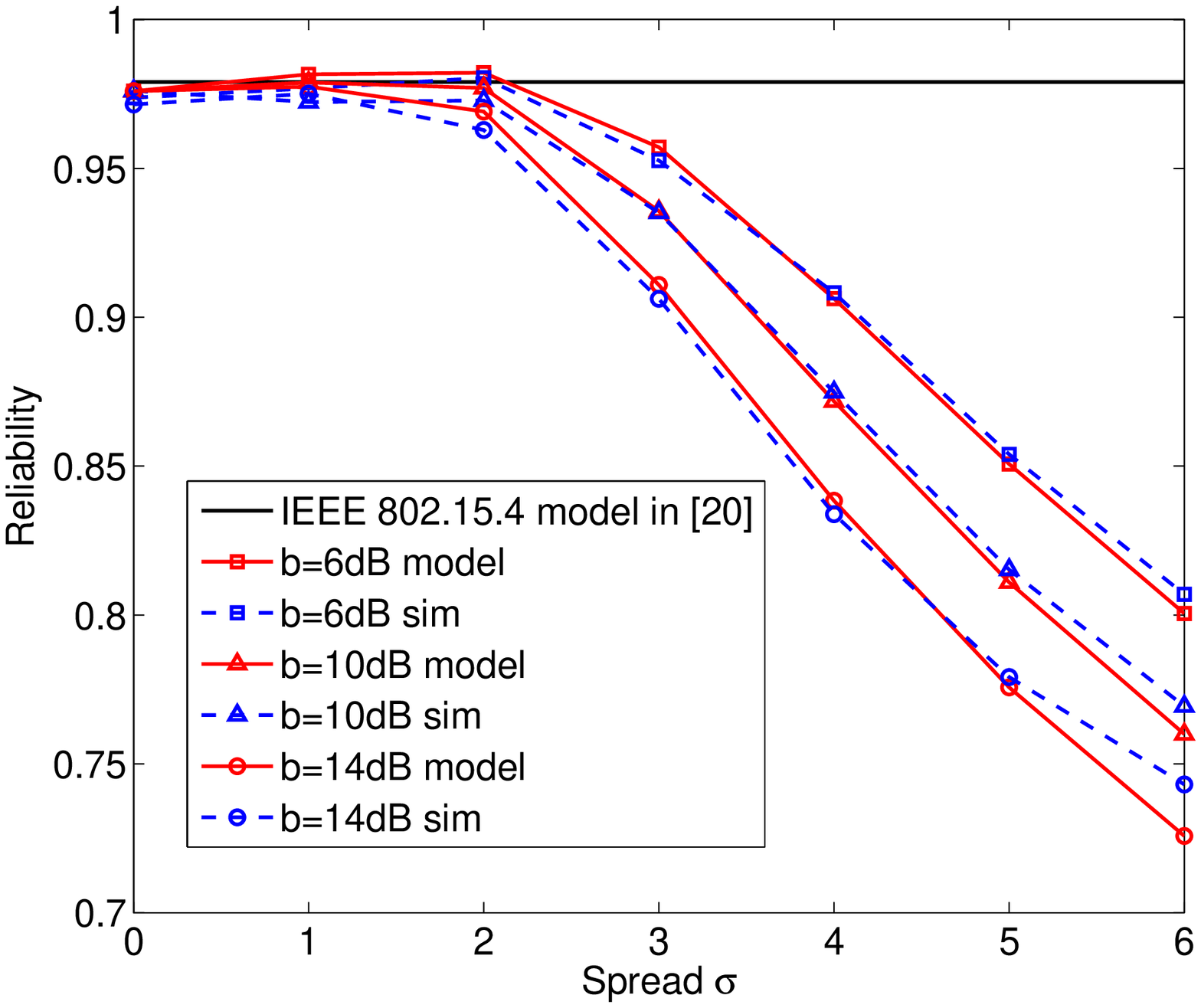}
\caption{Reliability vs. $\sigma$ for the star network in Fig.~\ref{fig:topology}a)  with $N=7$ nodes,  $r=1$~m, $\lambda=10$ pkt/s, $a=-76$~dB.}\label{fig:rel_sigma_b}\end{figure}

\begin{figure}[t] \centering
\includegraphics[width=0.59\textwidth]{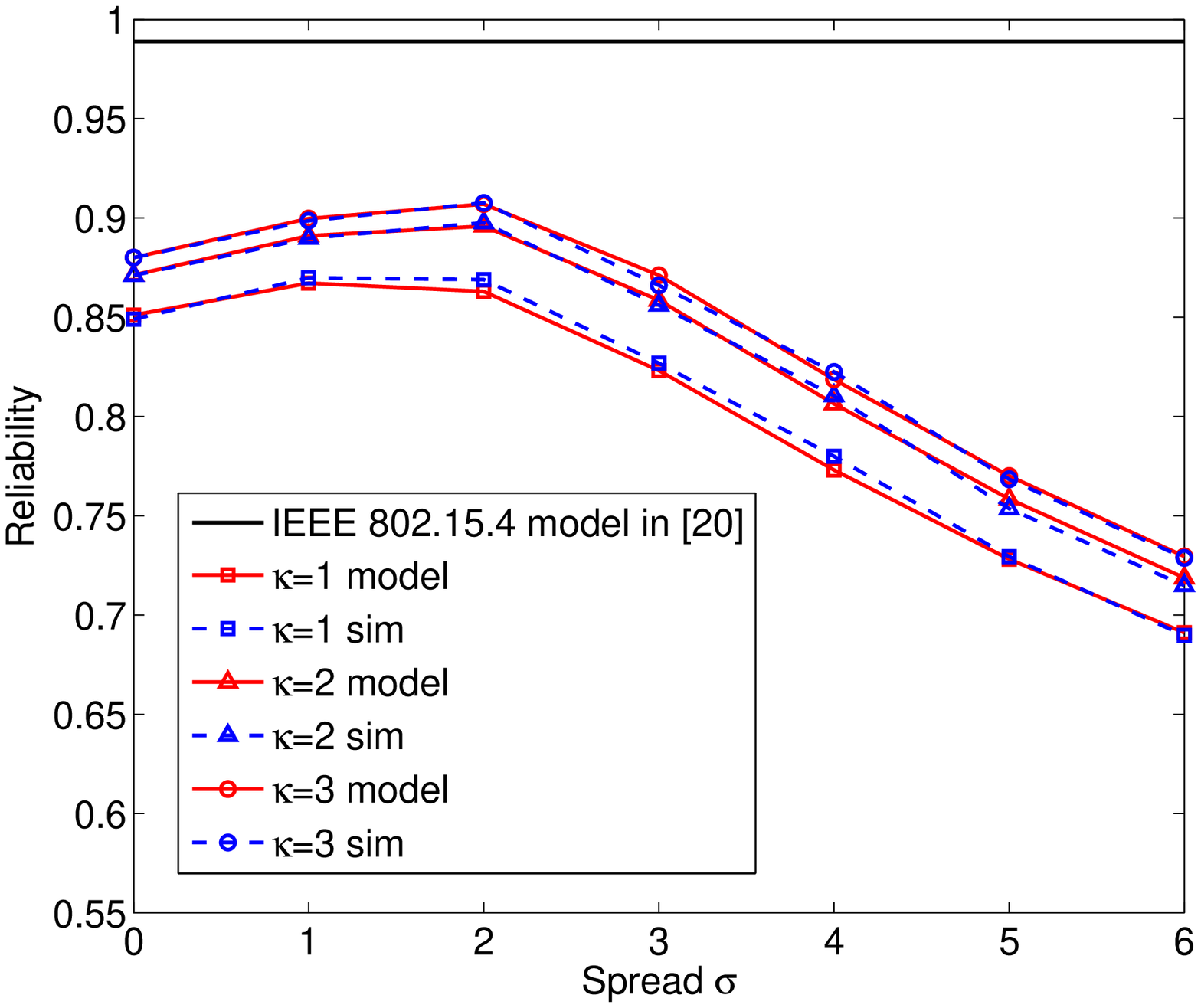}
\caption{Reliability vs. $\sigma$ for the star network in Fig.~\ref{fig:topology}a)  with $N=7$ nodes,  $r=1$~m, $\lambda=5$ pkt/s, $a=-56$~dB,  $b=6$~dB.}\label{fig:rel_sigma_k}\end{figure}

In this set of performance results, we consider a single-hop star topology as in
Fig.~\ref{fig:topology}a). We let the number of nodes be
$N=7$, the MAC parameters $m_0=3$, $m=4$, $m_b=5$, $n=0$, $L=70$~bytes,
$L_{\rm ack}=11$~bytes and the physical layer parameters
$P_{\mathrm{tx},i}=0$~dBm, and $k=2$. We validate our model and study
the performance of the network by varying the traffic rate $\lambda_i=\lambda$, $i=1,...,N$, in the range
$0.1 - 10$~pkt/s, the radius $r_{i,0}=r$, $i=1,...,N$, in the range $0.1 - 10$~m,
the spread of the shadow fading $\sigma_i=\sigma$, $i=1,...,N$, in the range $0 - 6$, and the Nakagami parameter $\kappa$ in the range $1 - 3$. 
The IEEE 802.15.4 standard specifies that the carrier sensing threshold is $10$~dB above the maximum receiver sensitivity for the physical layer (which is typically around $-85$~dBm)~\cite{ieee802154}.
Therefore, we show results for different values of the carrier sensing threshold, namely $a=-76$~dBm, $a=-66$~dBm, and $a=-56$~dBm.
The outage threshold is not specified by the standard. Experimental measurements for IEEE 802.15.4 show that the minimum SINR that guarantees correct packet reception is about $6$~dB~\cite{buratti_capture}.
In the following, we show results for different values of the outage threshold, namely, $b=6$~dB, $b=10$~dB, and $b=14$~dB.

In Fig.~\ref{fig:rel_lambda}, we report the average reliability
over all links by varying the node traffic rate $\lambda$. 
The results
are shown for different values of the spread $\sigma$ and in the absence of multi-path ($f_i=1$). 
The model is compared with the results obtained by using the model in~\cite{PG_TVT},
which was developed in the absence of a channel model.
There is a good matching between the simulations and the analytical expression~\eqref{eq:reliability}.
The reliability decreases as the traffic increases. Indeed, an increase of the traffic generates an
increase of the contention level at MAC layer. 
Our model is close to the ideal case in~\cite{PG_TVT} in the absence of stochastic fluctuation of the
channel ($\sigma$=0). The small gap is due to the presence of thresholds for channel sensing and outage,
which reduce the reliability due to possible failures in the CCA mechanism. 
However, a remarkable aspect is that the impact
of shadow fading is more relevant than variations in the traffic.
Therefore, a prediction based only on Markov chain analysis of the MAC without including the channel behavior,
as typically done in the previous literature, is largely inaccurate to capture the performance of IEEE~802.15.4 wireless networks, especially at larger shadowing spreads.

In Fig.~\ref{fig:del_lambda}, the average delay over all links
is reported. Also in this case simulation results follow quite well
results obtained from the model as given by Eq.~\eqref{eq:average_delay}. 
The delay in our model with $\sigma$=0 is lower than the delay evaluated in the model in~\cite{PG_TVT}
due to the effects of thresholds for channel sensing and outage,
which reduce the reliability due to possible failures in the CCA mechanism.
An increase of traffic leads to an increase of the average delay due to the
larger number of channel contentions and consequently an increase in the number of
backoffs. The spread of shadowing components does not impact on the delay
significantly, particularly for low traffic, because lost packets
due to fading are not accounted for in the delay computation.
When the traffic increases, we note that fading is actually
beneficial for the delay. In fact, the delay of successfully
received packets reduces by increasing $\sigma$. This is because
the occurrence of a deep fading reduces the probability of
successful transmission. However, since this holds for all nodes,
the average number of contending nodes for the CCA may reduce, thus reducing the average delay of
successfully received packets. It is not possible to capture this network behavior by using separate models of the IEEE~802.15.4 MAC and physical layers as in the previous literature, since this effect clearly depends on a cross-layer interaction.

In Fig.~\ref{fig:ene_lambda}, the
average power consumption over all links is presented and compared with the analytical expression in Eq.~\eqref{eq:energy}. The
number of packet transmissions and ACK receptions is the major
source of energy expenditure in the network. Therefore, an
increase of the traffic leads to an increase of the power
consumption, while performance are marginally affected by the
spread of the fading. However, the power consumption is slightly
reduced when the spread is $\sigma=6$, due to the smaller number of
received ACKs. Note that no power control policy is implemented.

In Fig.~\ref{fig:rel_distance}, the average reliability is
reported as a function of the radius $r$ for different values of the
spread $\sigma$. Again, analytical results obtained through Eq.~\eqref{eq:reliability} are in good agreement with those provided by simulations.
For the ideal channel case (i.e.,
$\sigma=0$) the size of the network does not affect the
reliability in the range $r=0.1 - 10$~m. For $\sigma=6$, the
performance degrades significantly as the radius increases. An
intermediate behavior is obtained for $\sigma=3$, where the
reliability is comparable to the ideal channel case for short links, but
 it reduces drastically for $r>1$~m. The effect is the
combination of an increase of the outage probability with the
radius (due to the path loss component) and hidden terminals that
are not detected by the CCA. 

In Fig.~\ref{fig:del_distance}, we
report the average delay by varying the radius $r$ for different
values of the spread $\sigma$. The shadowing affects the delay
positively and the effect is more significant for larger inter-node
distances: in this case the average
number of contending nodes for the free channel assessment
reduces, thus the busy channel probability reduces, which in turn
decreases the average delay of successfully received packets. 

In Fig.~\ref{fig:ene_distance}, the average power consumption by
varying $r$ is presented. We notice a similar behavior as for the
delay. The power consumption reduces with the fading and the
increasing size of the network. Nodes spend less time in the
backoff and channel sensing procedure due to reduced number of
contending nodes and the number of ACKs.

Fig.~\ref{fig:rel_sigma_a} shows the
average reliability as a function of the shadowing spread $\sigma$.
 The results are
plotted for different values of the carrier sensing threshold $a$.
 The reliability decreases when the
threshold $a$ become larger. The impact of the variation of
the threshold $a$ is maximum for $\sigma=0$, and the gap reduces
when the spread $\sigma$ increases.  In Fig.~\ref{fig:del_sigma_a}, the average
delay is plotted as a function of the spread $\sigma$. Depending on the
threshold $a$, the delay shows a different behavior when
increasing $\sigma$: it increases for $a=-76$~dBm and it decreases
for $a=-66$~dBm, and $a=-56$~dBm. As we discussed above, the
spread $\sigma$ may reduce the delay under some circumstances. However, when the
threshold is large, the average number of contenders is less
influenced by the fading and does not decrease significantly,
while the busy channel probability becomes dominant and the number
of backoffs increases, so that the delay increases as well.
Fig.~\ref{fig:ene_sigma_a} reports the average power consumption
by varying the spread $\sigma$. The power consumption reduces by
increasing the threshold $a$ as a consequence of the smaller number of
ACK transmissions, although a maximum consumption is observed for low values of the spread.

In Fig.~\ref{fig:rel_sigma_b}, we plot  the
average reliability as a function of the spread $\sigma$ for different values of the
outage threshold $b$.  The
threshold $b$ does not affect the performance noticeably for
$\sigma=0$, while the gap in the reliability increases with
$\sigma$. Note that for a high threshold the reliability tends to
increase with $\sigma$ as long as $\sigma$ is small or moderate, and it decreases for large
spreads. In our setup, a maximum in the reliability is obtained for
$\sigma \approx 2$.

In Fig.~\ref{fig:rel_sigma_k}, we report the combined effects of shadow fading and multi-path fading on the reliability.
We show the reliability as a function of the spread $\sigma$ of the
shadow fading for different values of the Nakagami parameter $\kappa$.
We recall that $\kappa=1$ corresponds to Rayleigh fading.
There is a good match between the simulations and the analytical model~\eqref{eq:reliability}.
The effect of the multi-path is a further degradation of the reliability. However,
the impact reduces as the
Nakagami parameter $\kappa$ increases and the fading becomes less severe. In fact, for $\kappa \gg 1$, the effect of multi-path
becomes negligible. Furthermore, the multi-path fading and the composite channel evidences the presence of the maximum at $\sigma \approx 2$ in the plot of reliability.

\subsection{Multi-hop Linear Topologies}

In this set of performance results, we consider the multi-hop linear topology in
Fig.~\ref{fig:topology}b). The number of nodes is $N=5$, with the same MAC and physical
layer parameters as in the single-hop case. We validate our model and study the performance
of the network as a function of the hop distance $r_{i,j}$ in the range
$r=0.1 - 10$~m, and the spread of the shadow fading in the range
$\sigma=0 - 6$. We show results for each hop, and for different
values of the carrier sensing threshold $a=-76, 66, 56$~dBm, and
outage threshold $b=6, 10, 14$~dB.

\begin{figure}[t] \centering
\includegraphics[width=0.59\textwidth]{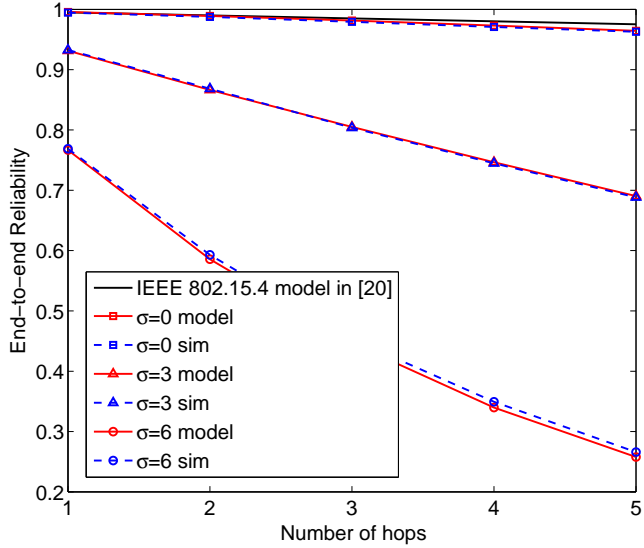}
\caption{End-to-end reliability vs. number of hops for the linear topology in Fig.~\ref{fig:topology}b) with $N=5$ nodes, $r=1$~m, $\lambda=2$~pkt/s, $a=-76$~dB, $b=6$~dB.}\label{fig:rel_linear_s}\end{figure}

\begin{figure}[t] \centering
\includegraphics[width=0.59\textwidth]{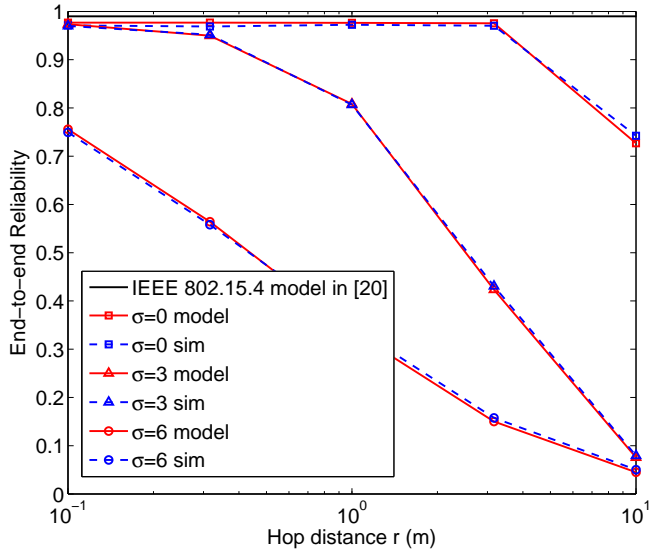}
\caption{End-to-end reliability vs. hop distance $r$ for  the linear topology in Fig.~\ref{fig:topology}b) with $N=5$ nodes, $\lambda=2$~pkt/s, $a=-76$~dB, $b=6$~dB.}\label{fig:rel_linear_r}\end{figure}

\begin{figure}[t] \centering
\includegraphics[width=0.59\textwidth]{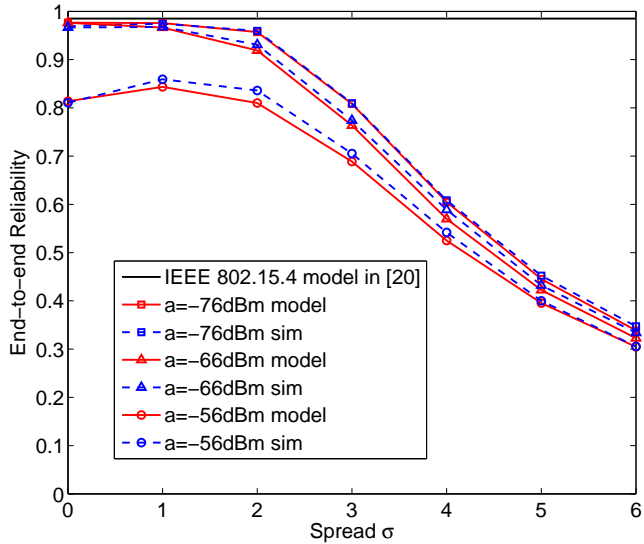}
\caption{End-to-end reliability vs. $\sigma$ for  the linear topology in Fig.~\ref{fig:topology}b) with $N=5$ nodes, $r=1$~m, $\lambda=2$~pkt/s, $b=6$~dB.}\label{fig:rel_linear_a}\end{figure}

\begin{figure}[t] \centering
\includegraphics[width=0.59\textwidth]{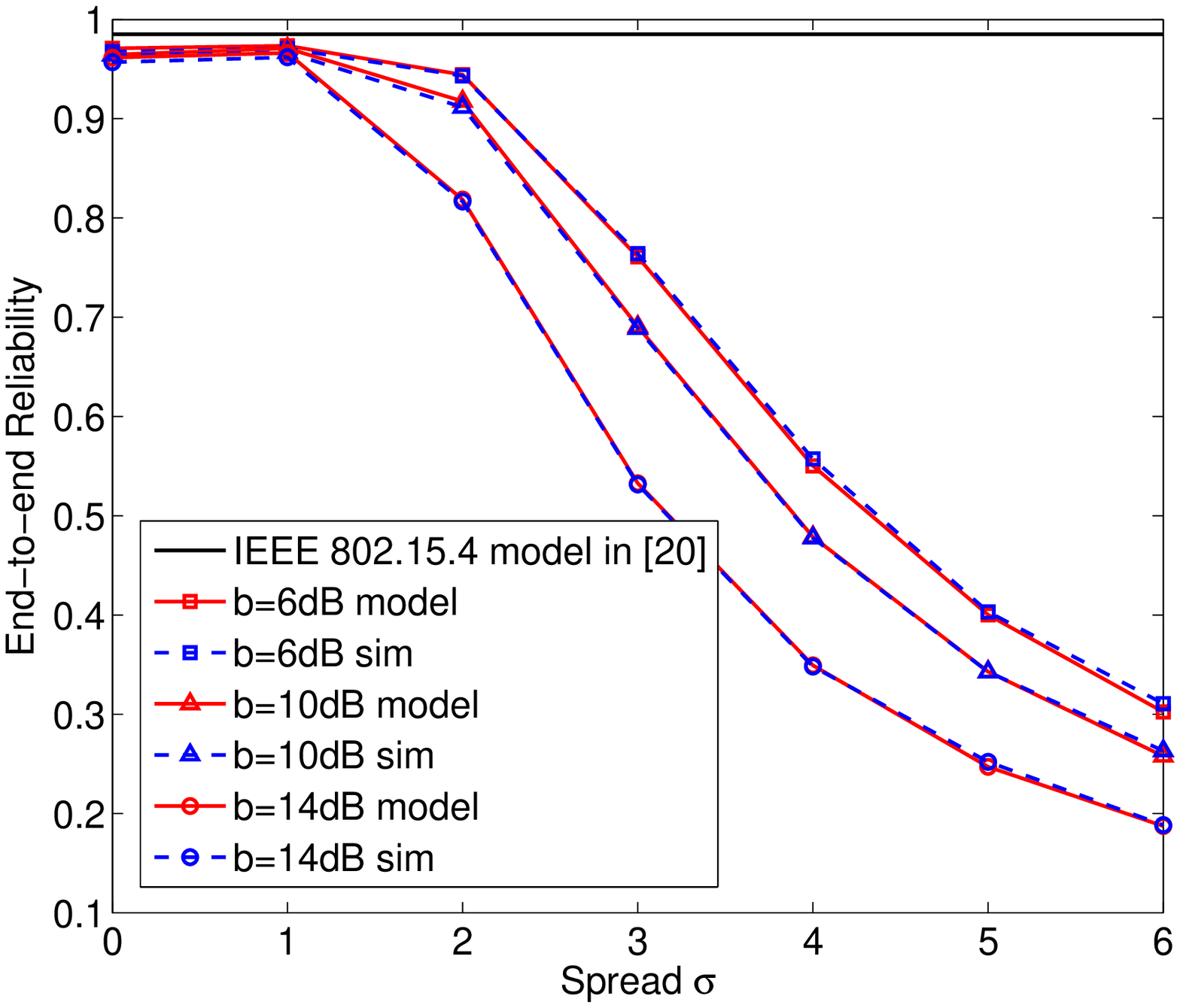}
\caption{End-to-end reliability vs. $\sigma$ for  the linear topology in Fig.~\ref{fig:topology}b) with  $r=1$~m, $\lambda=2$~pkt/s, $a=-76$~dB.}\label{fig:rel_linear_b}\end{figure}

In Fig.~\ref{fig:rel_linear_s}, the end-to-end reliability is
reported from each node to the destination node for different
values of the spread $\sigma$. The analytical model follows well
the simulation results. The end-to-end reliability decreases with
the number of hops. This effect is more evident in the presence of 
shadowing. Fig.~\ref{fig:rel_linear_r} shows the end-to-end
reliability from the farthest node to the destination by varying
the distance $r$ between every two adjacent nodes for different values
of the spread $\sigma$. The reliability is very sensitive to an
increase of the hop distance. In
Fig.~\ref{fig:rel_linear_a}, we show the end-to-end reliability by
varying the spread $\sigma$ of the shadow fading. Results are
shown for different values of the carrier sensing threshold $a$. In
Fig.~\ref{fig:rel_linear_b}, we plot the end-to-end reliability
for different values of $b$. Similar considerations as for the
single-hop case applies here. However, for the linear topology,
the reduction of the carrier sensing range from $a=-76$~dBm to
$a=-66$~dBm influences less the reliability since hidden nodes are
often out of range of the receiver, 
therefore the channel detection failure may not lead to
collisions.

\subsection{Multi-hop Topologies with Multiple End-devices}

%


\begin{figure}[t] \centering
\includegraphics[width=0.59\textwidth]{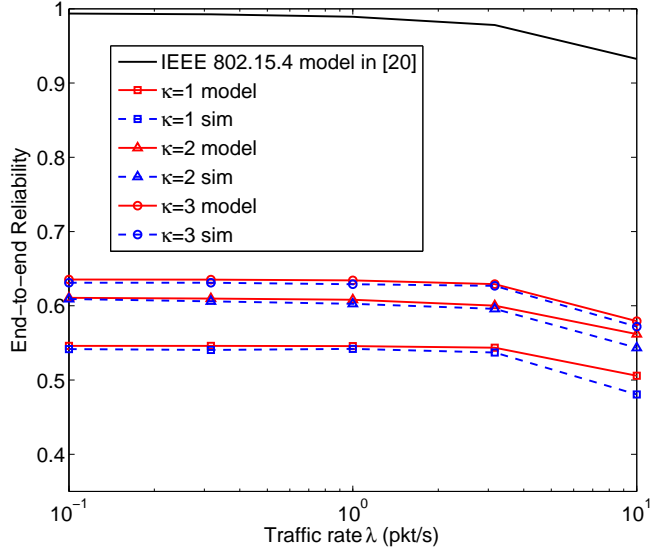}
\caption{End-to-end reliability vs. traffic rate $\lambda$ for the multi-hop topology in Fig.~\ref{fig:topology}c) with  $a=-76$~dB, $b=6$~dB, $\sigma=6$.}\label{fig:rel_gen}\end{figure}

\begin{figure}[t] \centering
\includegraphics[width=0.59\textwidth]{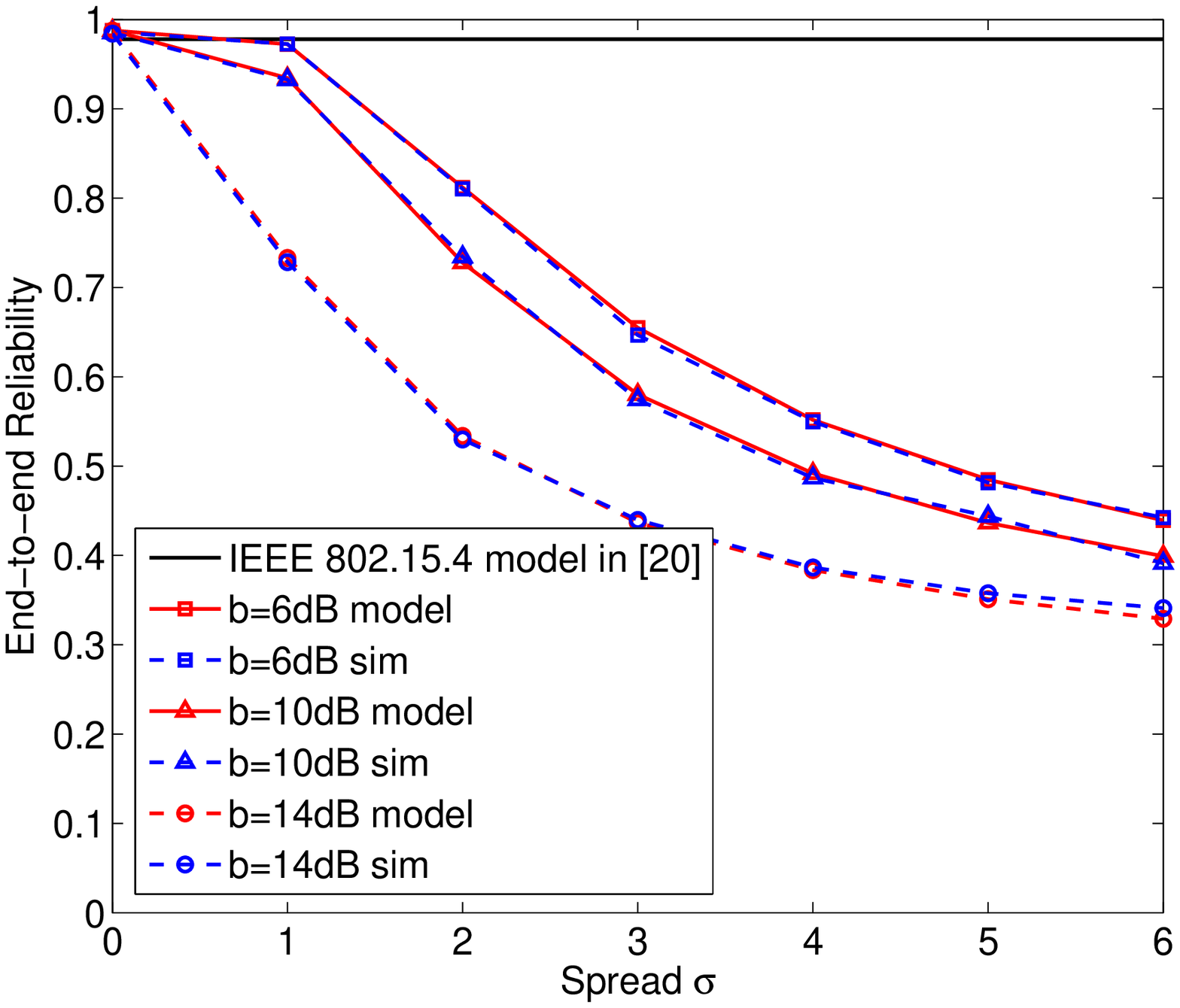}
\caption{End-to-end reliability vs. $\sigma$ for the multi-hop topology in Fig.~\ref{fig:topology}c)  with $\lambda=2$~pkt/s, $a=-76$~dB.}\label{fig:rel_gen_b}\end{figure}

We consider the multi-hop topology in
Fig.~\ref{fig:topology}c). We use the same MAC and physical
layer parameters as in the single-hop case.
We consider the end-to-end reliability as the routing metric and study the performance
of the network as a function of the traffic $\lambda_i=\lambda$, $i=1,...,N$, in the range
$0.1 - 10$~pkt/s, the spread of the shadow fading in the range
$\sigma=0 - 6$. Moreover, we show results for different
values of the Nakagami parameter $\kappa=1 - 3$ and threshold $b=6, 10, 14$~dB.

In Fig.~\ref{fig:rel_gen}, we report the average end-to-end reliability
over all the end-devices by varying the node traffic rate. The results
are shown for different values of Nakagami parameter $\kappa$ with the shadowing spread set to $\sigma=6$.
The impact of the Nakagami parameter $\kappa$ seems more prominent than variation of the traffic.
Fig.~\ref{fig:rel_gen_b} shows the end-to-end reliability by varying the spread $\sigma$
for different values of $b$. Differently to the other topologies,
a variation of the outage threshold $b$ has a strong impact on the reliability also for small to moderate shadowing spread.
In fact, due to the variable distance between each source-destination pair,
the fading and the outage probabilities affect the network noticeably.
This effect is well predicted by the developed analytical model.

\section{Conclusions}\label{sec:conclusions}

In this paper, we proposed an integrated cross-layer model of the MAC and physical
layers for unslotted IEEE 802.15.4 networks, by considering
explicit effects of multi-path shadow fading channels and the presence of interferers. We studied the impact
of fading statistics on the MAC performance in terms of
reliability, delay, and power consumption, by varying traffic rates,
inter-nodes distances, carrier sensing range, and SINR threshold. We observed that the severity
of the fading and the physical layer thresholds have significant
and complex effects on all performance indicators, and the effects are well predicted by the new model.
In particular, the fading has a relevant negative impact on the reliability.
The effect is more evident as traffic and distance between nodes increase.
However, depending on the carrier sensing and SINR thresholds, our model shows that a fading with small spread can improve the reliability with respect to the ideal case. The delay for successfully received packets and the power consumption are instead positively affected by the fading and the performance can be optimized by properly tuning the thresholds.

We believe that the design of future WSN-based systems can greatly benefit from
the results presented in this paper. As a future work, a tradeoff
between reliability, delay, and power consumption can be exploited
by proper tuning of routing, MAC, and physical layer parameters.
Various routing metrics can be analyzed, and the model extended to multiple sinks. 

\bibliographystyle{IEEEtran}
\bibliography{ref}

\end{document}